\documentclass{article}
\usepackage[utf8]{inputenc}
\usepackage{gensymb}
\usepackage{graphicx}
\usepackage{amssymb}
\usepackage{amsthm}
\usepackage{subfigure,subfloat}
\usepackage{authblk}
\usepackage{amsmath}
\usepackage{url}
\usepackage{xcolor}
\usepackage{upgreek}

\usepackage{geometry}
\graphicspath{{figs/}}

\title{Fungal Automata}

\author[1]{Andrew Adamatzky}
\author[1,2]{Eric Goles}
\author[1,3]{Genaro J. Mart{\'i}nez}
\author[1]{Michail-Antisthenis Tsompanas}
\author[4]{Martin Tegelaar}
\author[4]{Han A. B. Wosten}

\affil[1]{Unconventional Computing Laboratory, UWE, Bristol, UK}
\affil[2]{Faculty of Engineering and Science, University of Adolfo Ib\'{a}\~{n}ez, Santiago, Chile}
\affil[3]{High School of Computer Science, National Polytechnic Institute, Mexico}
\affil[4]{Microbiology Department, University of Utrecht, Utrecht, The Netherlands}

\date{}

\begin{document}

\maketitle

\begin{abstract}
\noindent
We study a cellular automaton (CA) model of information dynamics on a single hypha of a fungal mycelium. Such a filament is divided in compartments (here also called cells) by septa. These septa are invaginations of the cell wall and their pores allow for flow of cytoplasm between compartments and hyphae. The septal pores of the fungal phylum of the Ascomycota can be closed by organelles called Woronin bodies. Septal closure is increased when the septa become older and when exposed to stress conditions. Thus, Woronin bodies act as informational flow valves. The one dimensional fungal automata is a binary state ternary neighbourhood CA, where every compartment follows one of the elementary cellular automata (ECA) rules if its pores are open and either remains in state `0' (first species of fungal automata) or its previous state (second species of fungal automata) if its pores are closed. The Woronin bodies closing the pores are also governed by ECA rules. We analyse a structure of the composition space of cell-state transition and pore-state transitions rules, complexity of fungal automata with just few Woronin bodies, and exemplify several important local events in the automaton dynamics.
 \\
 
\vspace{2mm}

\noindent
\emph{Keywords:} fungi, ascomycete, Woronin body, cellular automata
\end{abstract}

\section{Introduction}

The fungal kingdom represents organisms colonising all ecological niches ~\cite{carlile2001fungi} where they play a key role ~\cite{griffin1972ecology,cooke1984ecology,rayner1988fungal,christensen1989view}. Fungi can consist of a single cell, can form enormous underground networks ~\cite{smith1992fungus} and can form microscopic fruit bodies or fruit bodies weighting up to half a ton ~\cite{dai2011fomitiporia}. The underground mycelium network can be seen as a distributed communication and information processing system linking together trees, fungi and bacteria~\cite{bonfante2009plants}. Mechanisms and dynamics of information processing in mycelium networks form an unexplored field with just a handful of papers published related to space exploration by mycelium ~\cite{held2009fungal,held2008examining}, patterns of electrical activity of fungi ~\cite{slayman1976action,olsson1995action,adamatzkyspiking} and potential use of fungi as living electronic and computing devices~\cite{adamatzky2018towards, adamatzky2020boolean,beasley2020memristive}.

Filamentous fungi grow by means of hyphae that grow at their tip and that branch sub-apically. Hyphae may be coenocytic or divided in compartments by septa. Filamentous fungi in the phylum \emph{Ascomycota} have porous septa that allow for cytoplasmic streaming~\cite{moore1962fine,lew2005mass}. Woronin bodies plug the pores of these septa after hyphal wounding to prevent excessive bleeding of cytoplasm~\cite{trinci1974occlusion, collinge1985woronin,jedd2000new,tenney2000hex,soundararajan2004woronin, maruyama2005three}. In addition, they plug septa of intact growing hyphae to maintain intra- and inter-hyphal heterogeneity~\cite{bleichrodt2012hyphal,bleichrodt2015switching,steinberg2017woronin,steinberg2017woronin,tegelaar2020subpopulations}. 

\begin{figure}[!tbp]
    \centering
   \subfigure[]{ \includegraphics[scale=0.5]{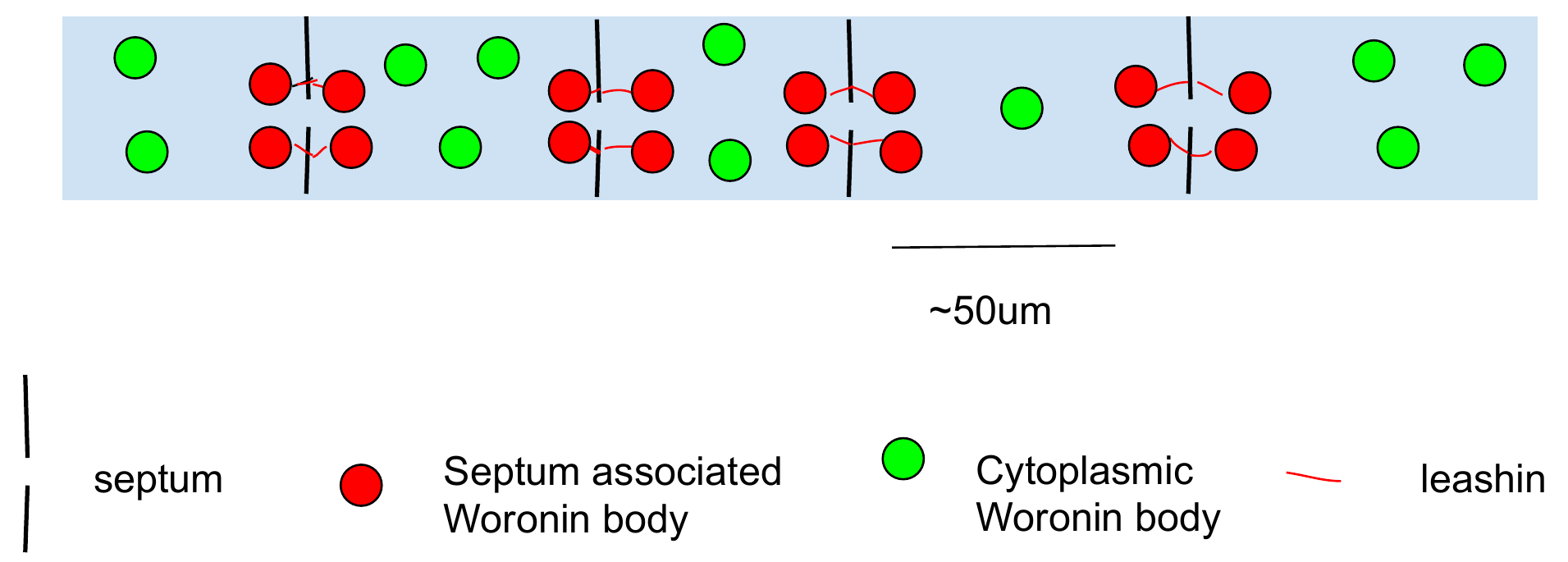}}
      \subfigure[]{ \includegraphics[scale=0.24]{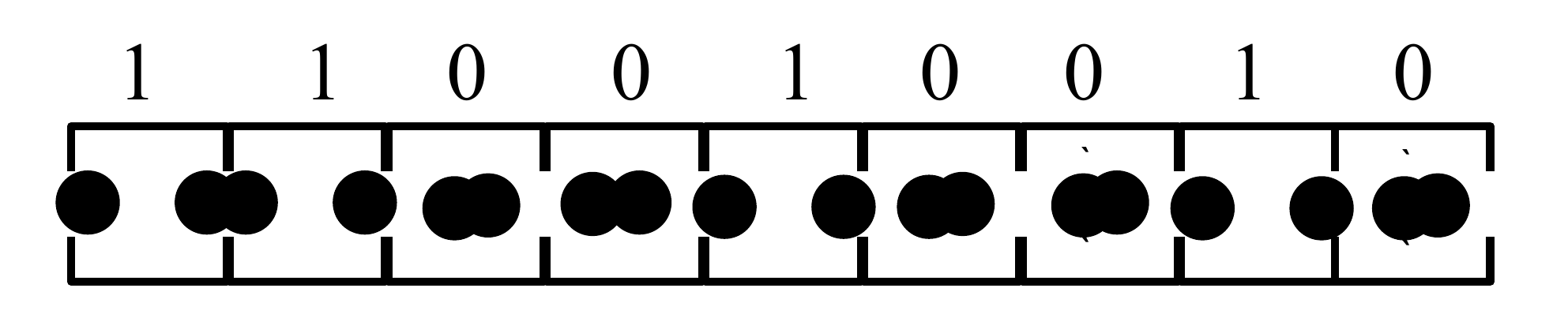}}
    \caption{(a)~A biological scheme of a fragment of a fungal hypha of an ascomycete, where we can see septa and associated Woronin bodies. (b)~A scheme representing states of Woronin bodies: `0' open, `1' closed.}
    \label{fig:bioscheme}
\end{figure}

Woronin bodies can be located in different hyphal positions  (Fig.~\ref{fig:bioscheme}a). When first formed, Woronin bodies are generally localised to the apex~\cite{momany2002mapping,tey2005polarized,beck2013characterization}. Subsequently, Woronin bodies are either transported to the cell cortex (\emph{Neurospora crassa}, \emph{Sordaria fimicola}) or to the septum (\emph{Aspergillus oryzae}, \emph{Aspergillus nidulans}, \emph{Aspergillus fumigatus}, \emph{Magnaporthe grisea}, \emph{Fusarium oxysporum}, \emph{Zymoseptoria tritici}) where they are anchored with a leashin tether and largely immobile until they are translocated to the septal pore due to cytoplasmic flow or ATP depletion \cite{ng2009tether,steinberg2017woronin,soundararajan2004woronin,momany2002mapping,maruyama2005three,tey2005polarized,wergin1973development,leonhardt2017lah,berns1992optical}. Woronin bodies that are not anchored at the cellular cortex or the septum, are located in the cytoplasm and are highly mobile (\emph{Aspergillus fumigatus}, \emph{Aspergillus nidulans}, \emph{Zymoseptoria tritici})~\cite{beck2013characterization,momany2002mapping,steinberg2017woronin}.
Septal pore occlusion can be induced by bulk cytoplasmic flow ~\cite{steinberg2017woronin} or developmental~\cite{bleichrodt2015selective} and environmental cues, like puncturing of the cell wall, high temperature, carbon and nitrogen starvation, high osmolarity and low pH. Interestingly, high environmental pH reduces the proportion of occluded apical septal pores~\cite{tegelaar2020subpopulations}.   

Aiming to lay a foundation of an emerging paradigm of fungal intelligence --- distributed sensing and information processing in living mycelium networks --- we decided to develop a formal model of mycelium and investigate a role of Woronin bodies in potential information dynamics in the mycelium. 

The paper is structured as follows. We introduce fungal automata in Sect.~\ref{automata}. Properties of the composition of cell state transition and Woronin body state transition functions are studied in Sect.~\ref{composition}. Complexity of space-time configuration of fungal automata, where just few cells have Woronin bodies is studied in Sect.~\ref{complexity}. Section~\ref{localevents} exemplifies local events, which could be useful for computation with fungal automata, happening in fungal automata with sparsely but regularly positioned cells with Woronin bodies. The paper concludes with Sect.~\ref{conclusion}.

\section{Fungal automata $\mathcal M$}
\label{automata}

\begin{figure}[!tbp]
    \centering
\subfigure[$\mathcal{M}_1, \rho_f=133,\rho_g=116$]{\includegraphics[width=0.26\textwidth]{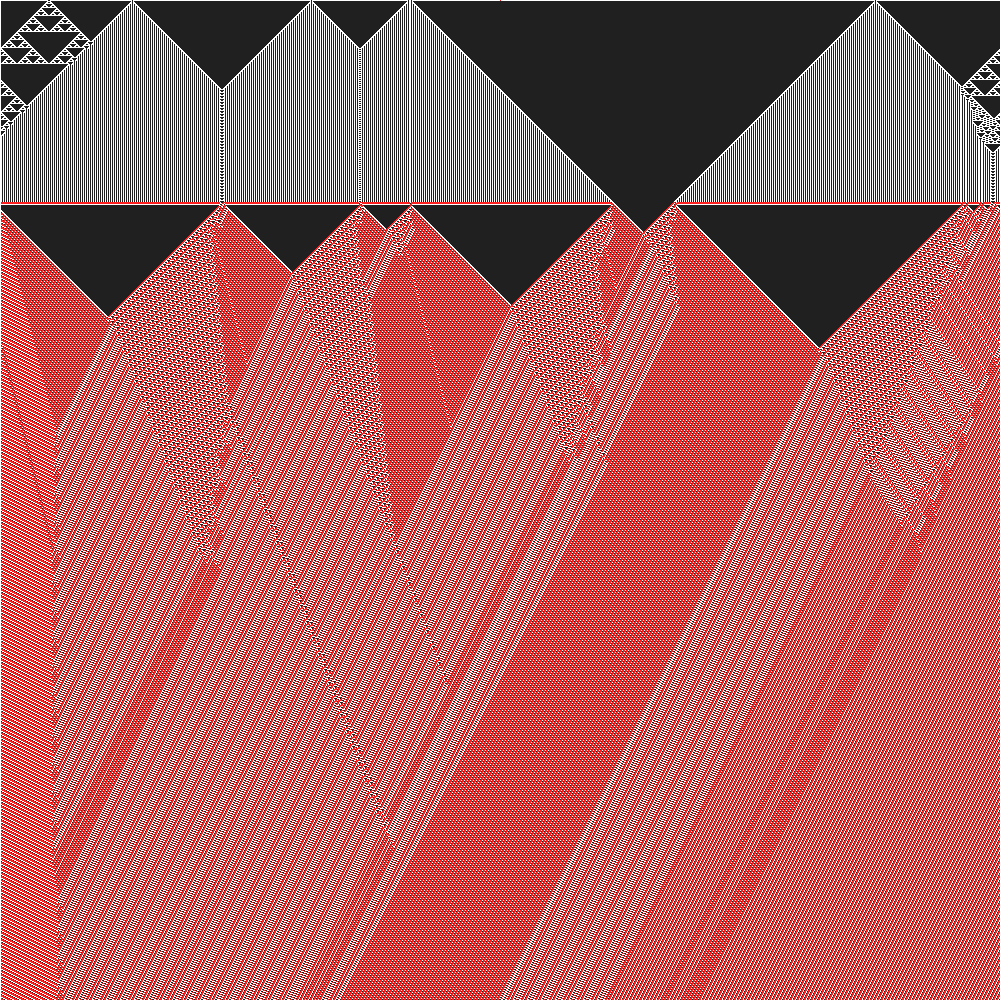}}
\subfigure[$\mathcal{M}_2, \rho_f=133,\rho_g=116$]{\includegraphics[width=0.26\textwidth]{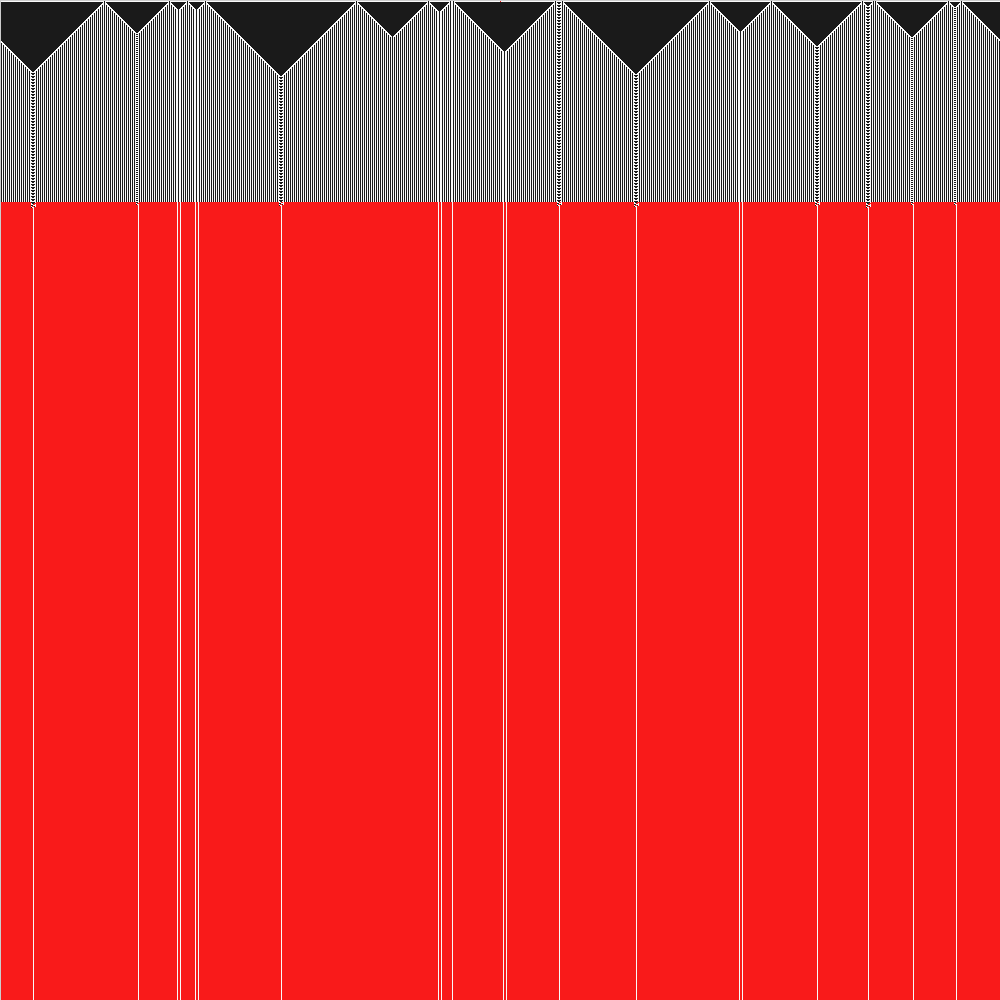}}
\subfigure[$\mathcal{M}_1, \rho_f=73,\rho_g=128$]{\includegraphics[width=0.26\textwidth]{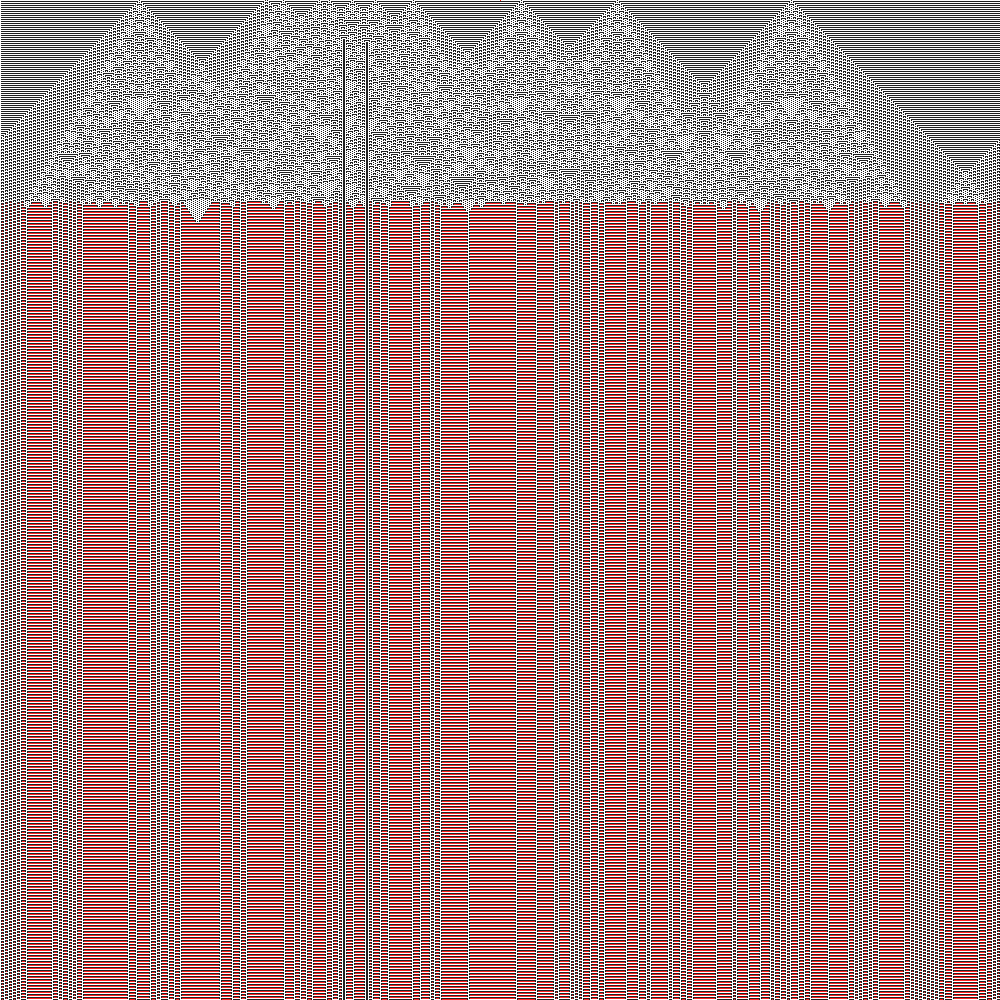}}
\subfigure[$\mathcal{M}_2, \rho_f=73,\rho_g=128$]{\includegraphics[width=0.26\textwidth]{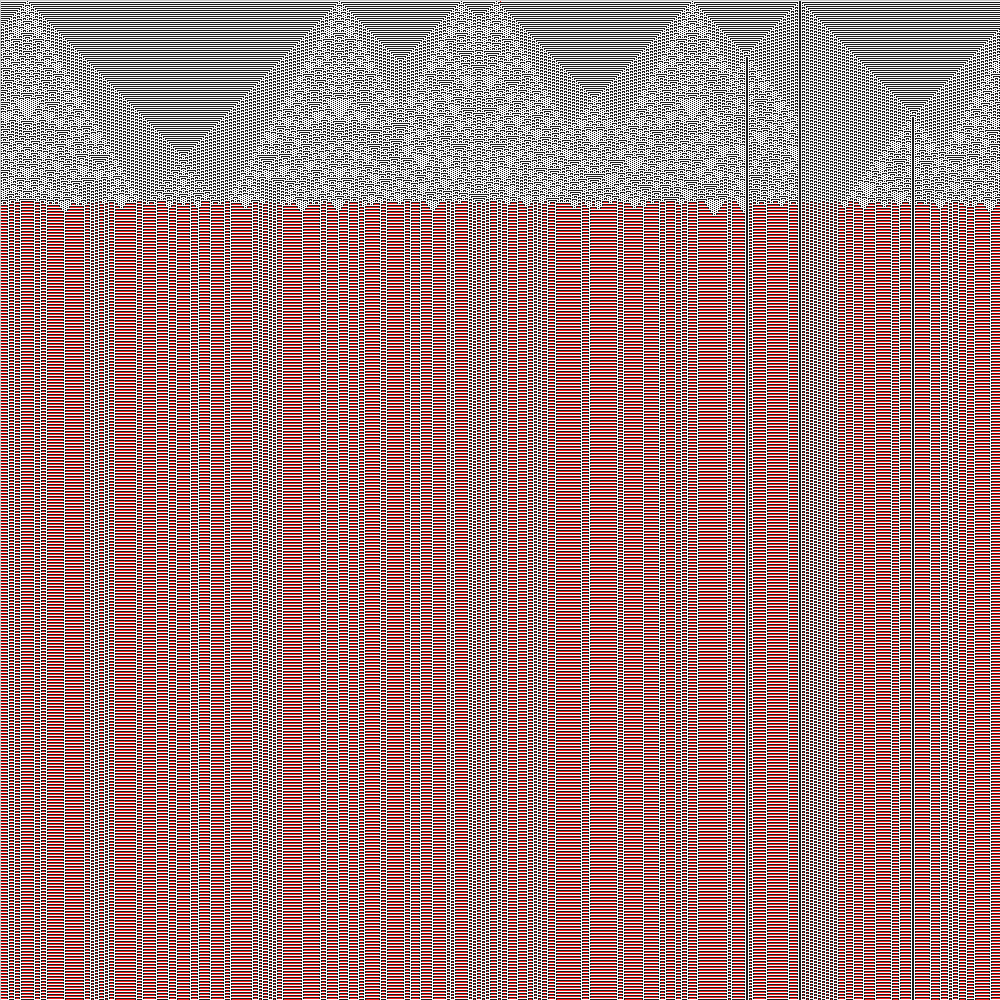}}
\subfigure[$\mathcal{M}_1, \rho_f=61,\rho_g=132$]{\includegraphics[width=0.26\textwidth]{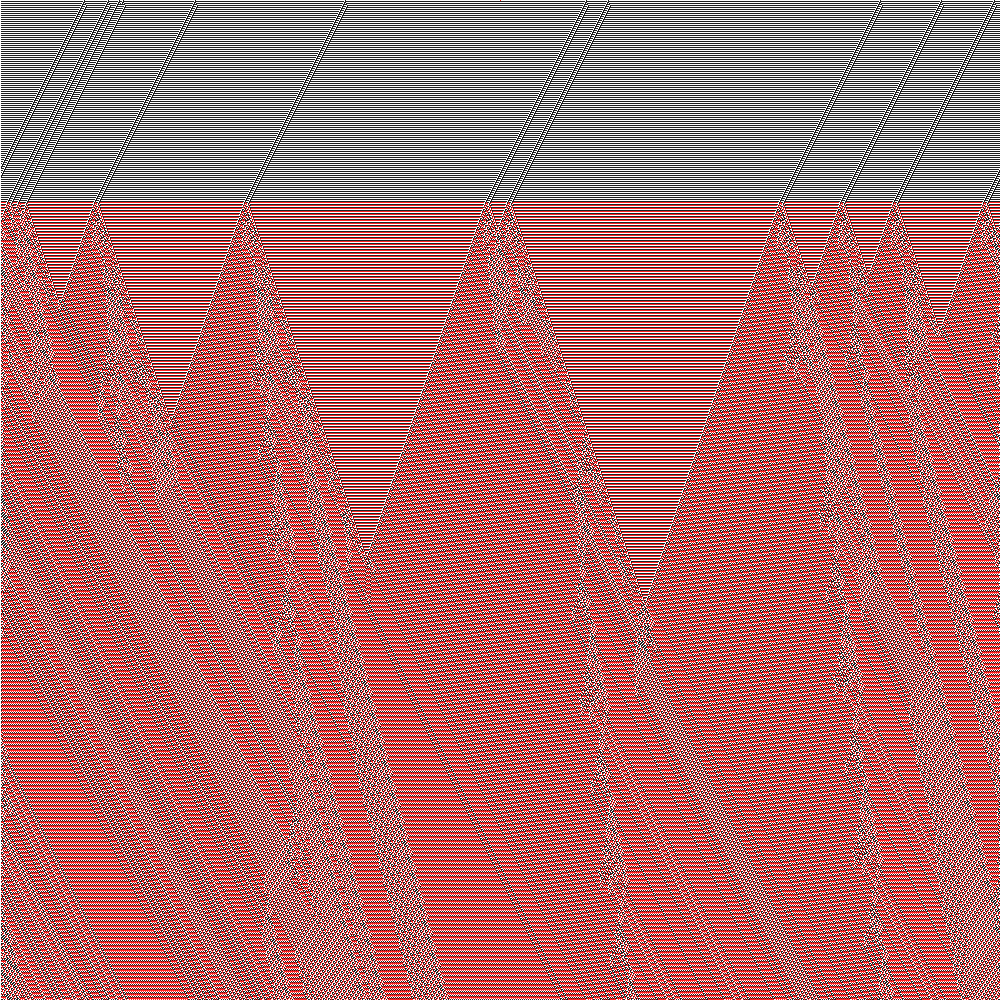}}
\subfigure[$\mathcal{M}_2, \rho_f=61,\rho_g=132$]{\includegraphics[width=0.26\textwidth]{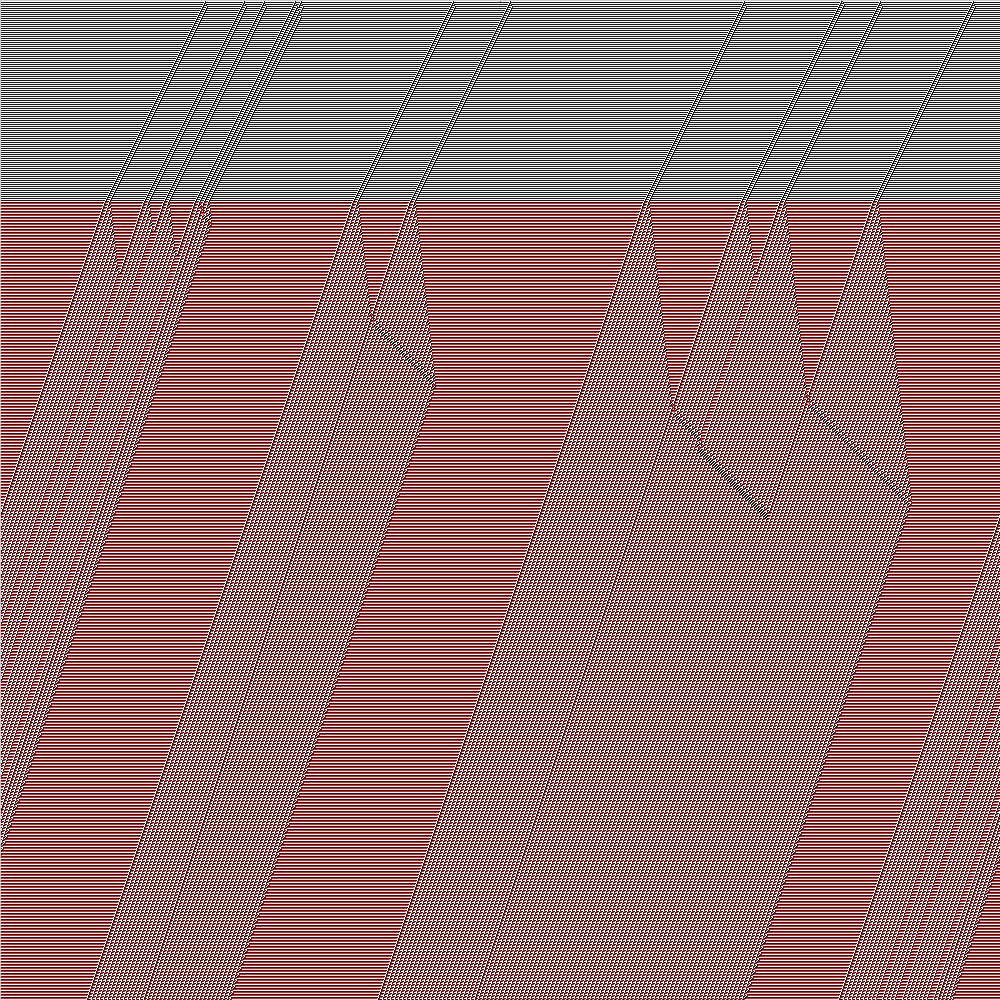}}
\subfigure[$\mathcal{M}_1, \rho_f=57,\rho_g=98$]{\includegraphics[width=0.26\textwidth]{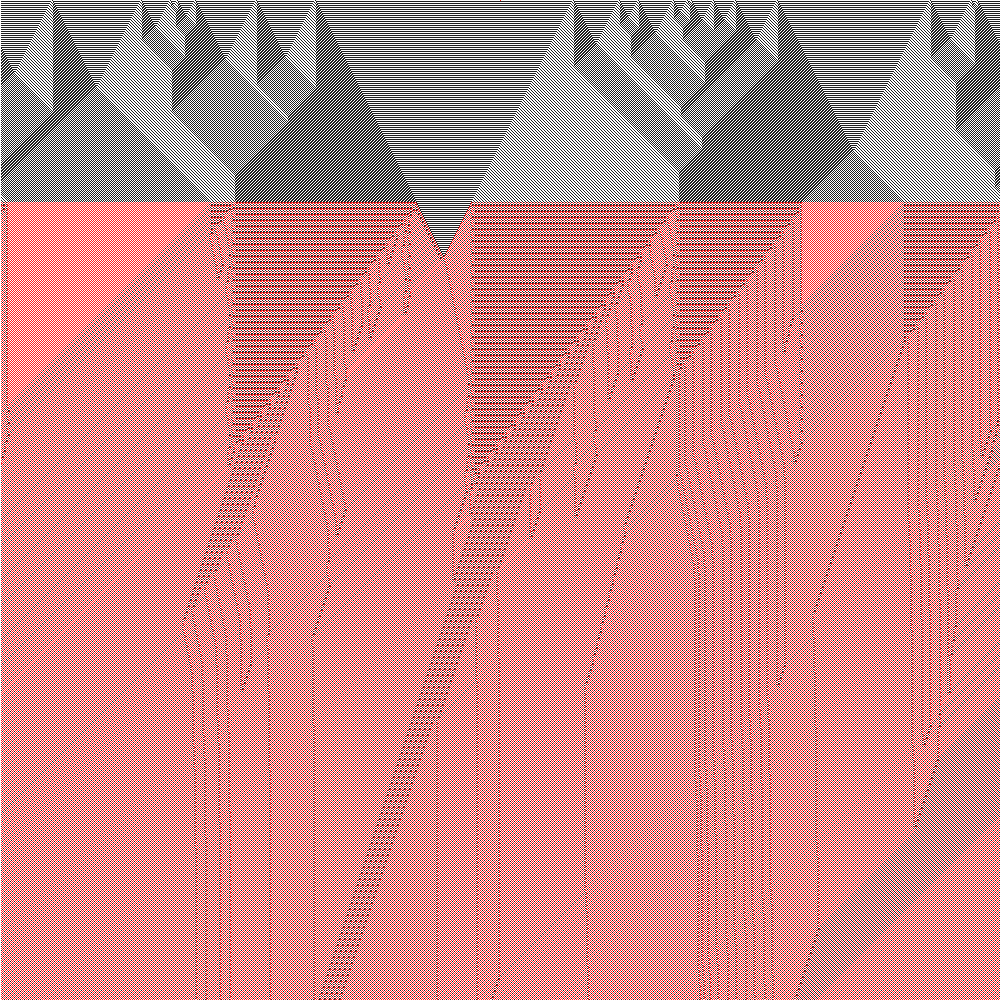}}
\subfigure[$\mathcal{M}_2, \rho_f=57,\rho_g=98$]{\includegraphics[width=0.26\textwidth]{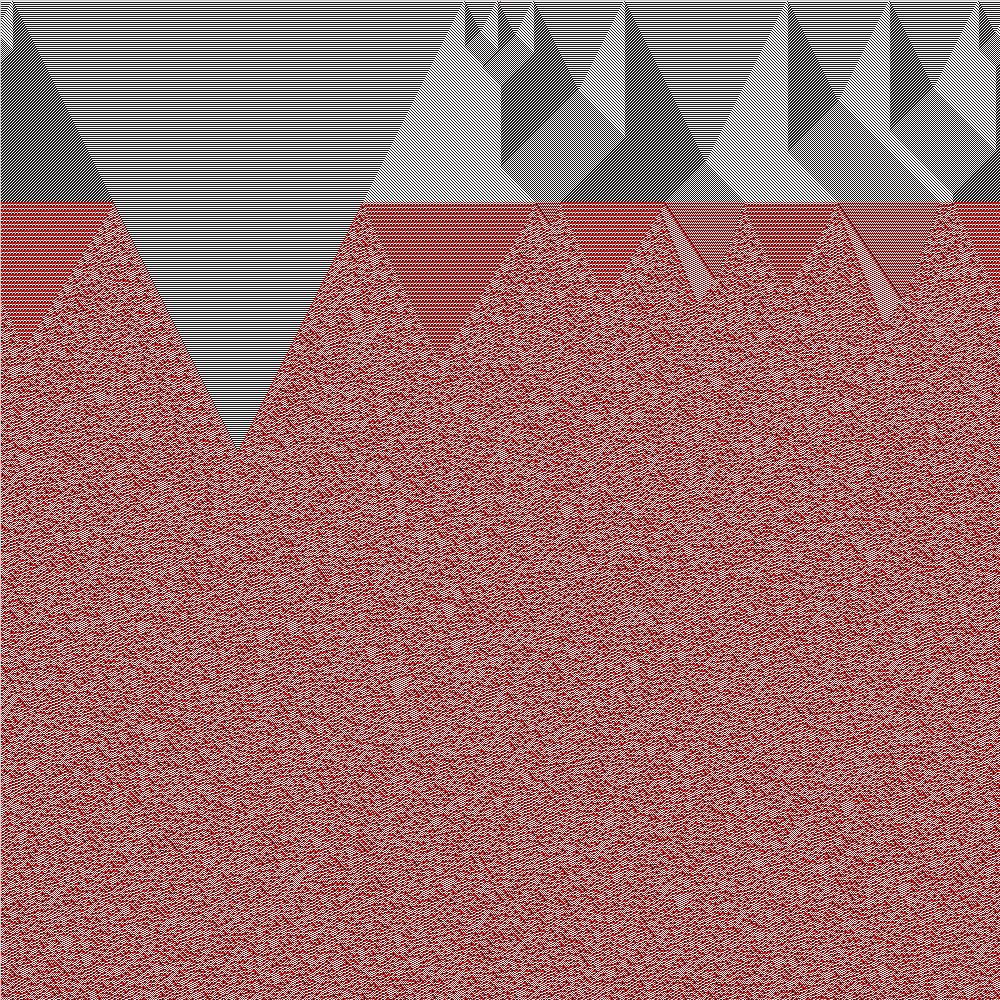}}
\subfigure[$\mathcal{M}_1, \rho_f=26,\rho_g=84$]{\includegraphics[width=0.26\textwidth]{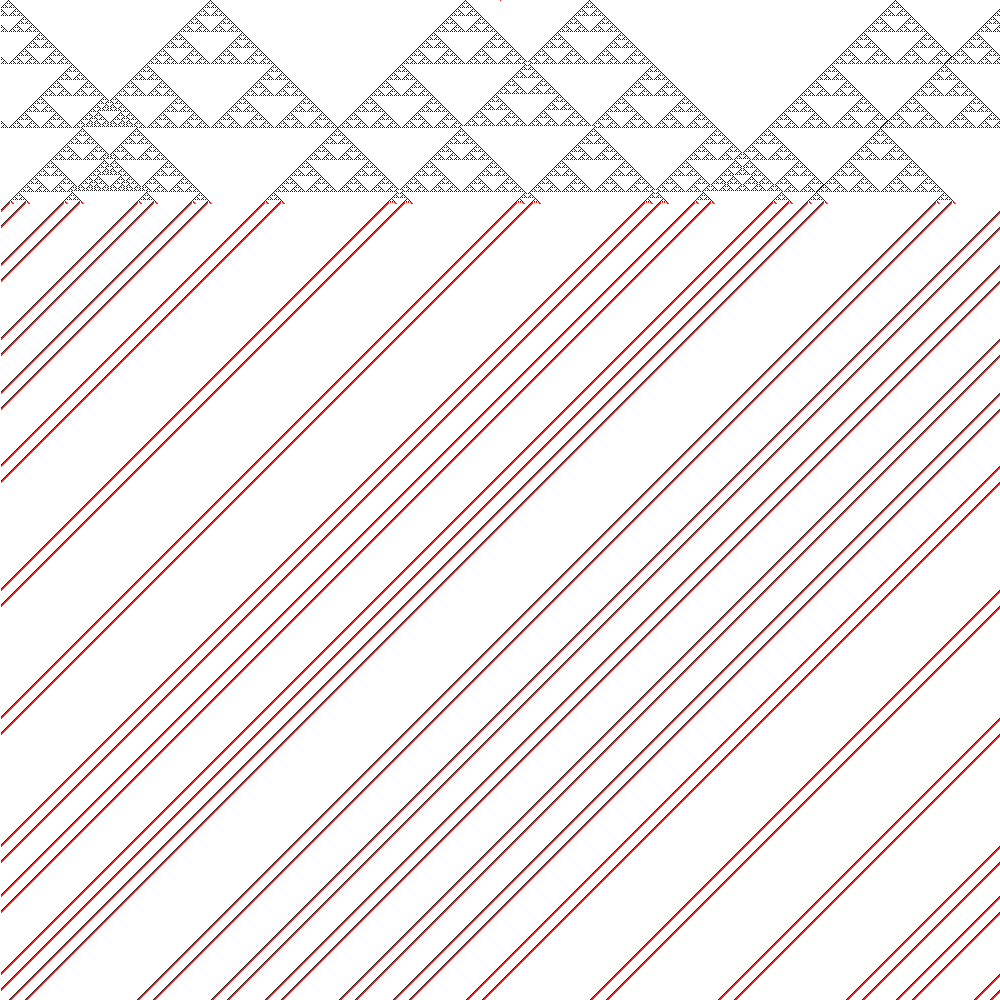}}
\subfigure[$\mathcal{M}_2, \rho_f=26,\rho_g=84$]{\includegraphics[width=0.26\textwidth]{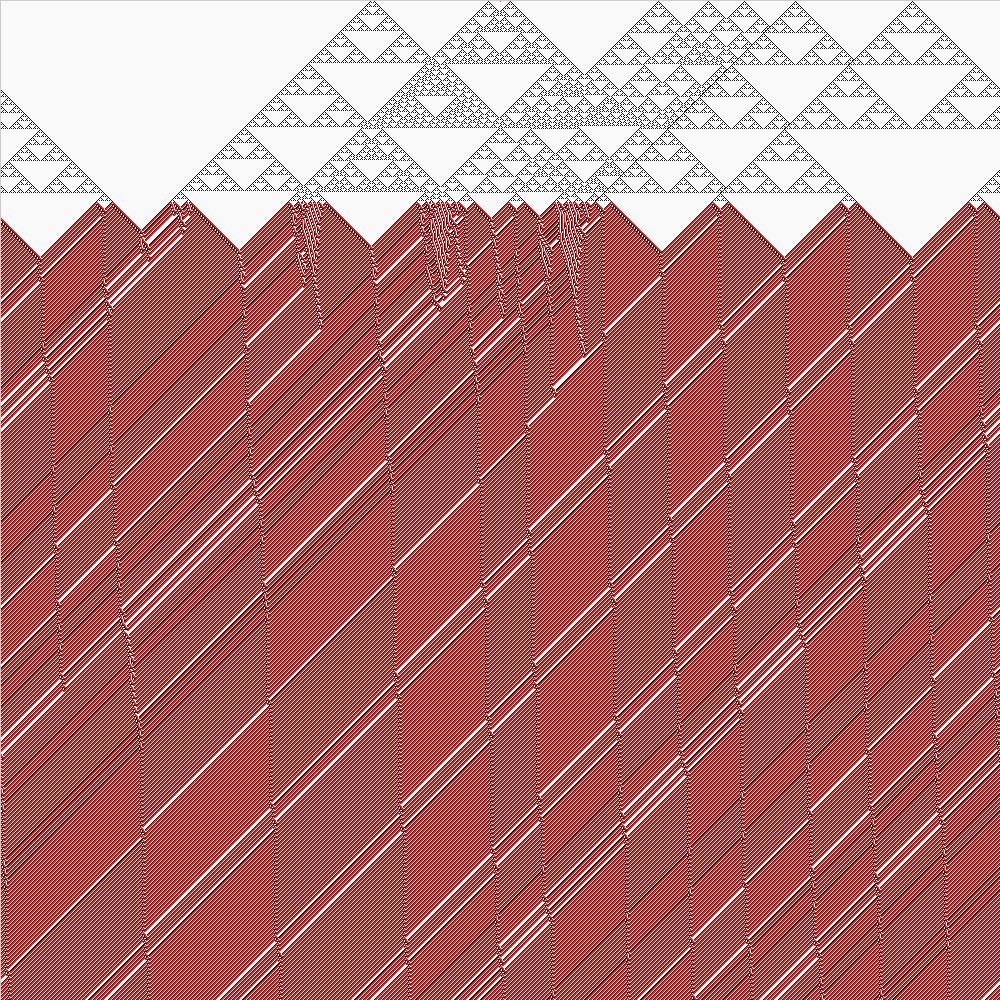}}
\subfigure[$\mathcal{M}_1, \rho_f=125,\rho_g=105$]{\includegraphics[width=0.26\textwidth]{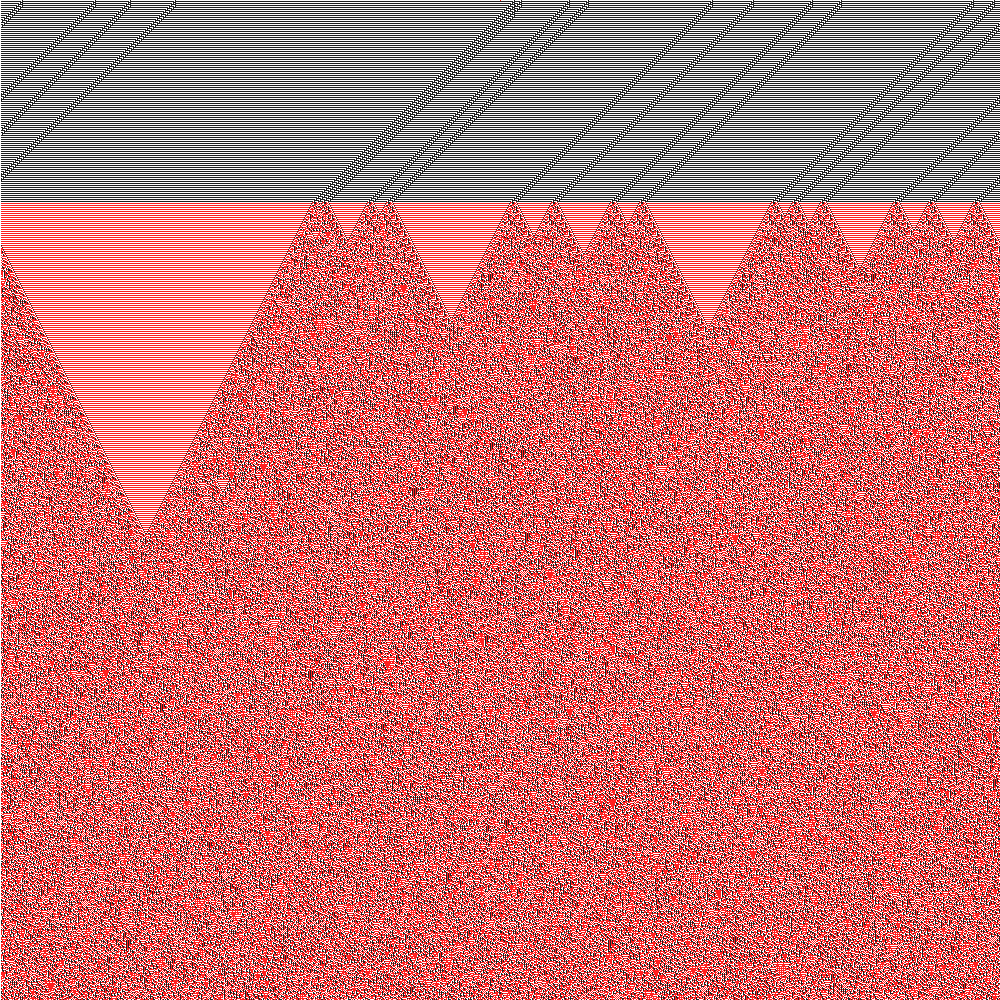}}
\subfigure[$\mathcal{M}_2,  \rho_f=125,\rho_g=105$]{\includegraphics[width=0.26\textwidth]{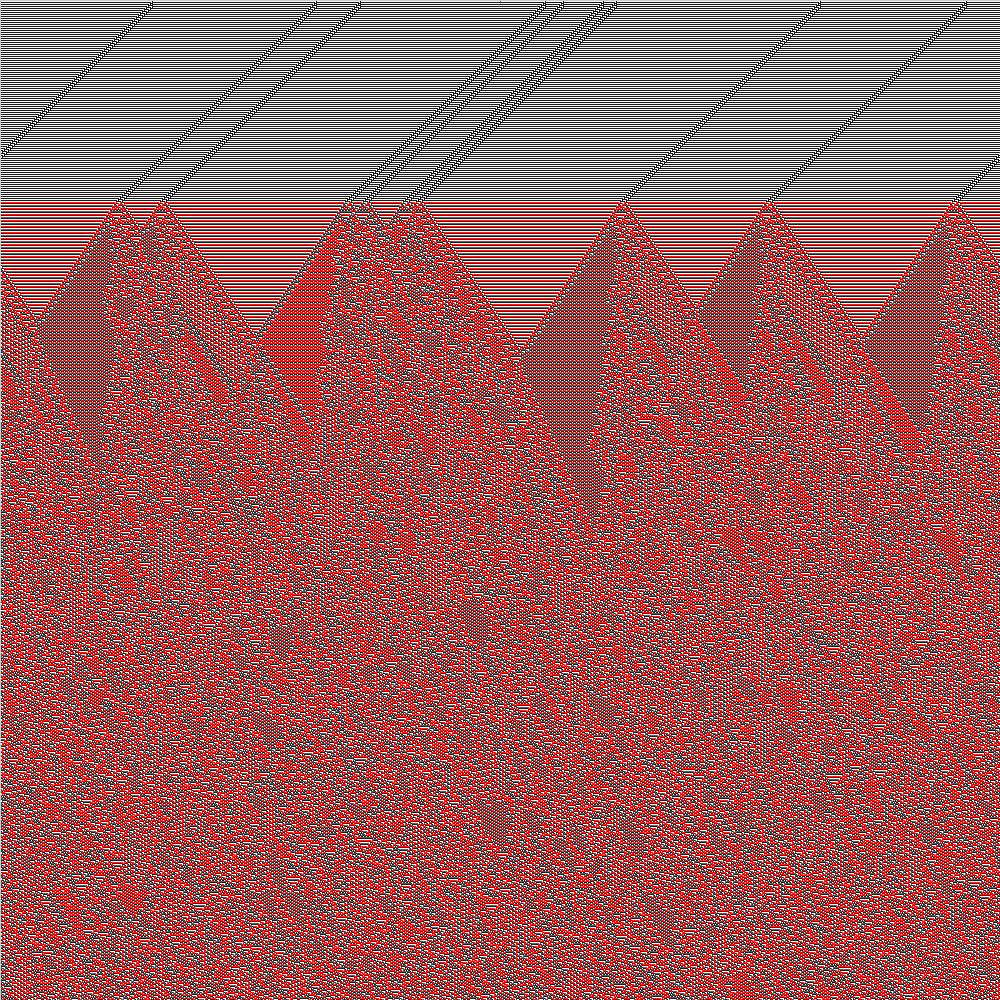}}
    \caption{Examples of space-time dynamics of $\mathcal M$. The automata are $10^3$ cells each. Initial configuration is random with probability of a cell $x$ to be in state `1', $x^0=1$, equals 0.01. Each automaton evolved for $10^3$ iterations. Binary values of ECA rules $f$ and $g$ are shown in sub-captions. Rule $g$ is applied to every iteration starting from 200th. Cells in state `0' are white, in state `1' are black, cells with Woronin bodies blocking pores are red. Indexes of cells increase from the left to the right, iterations are increasing from the to the bottom.}
    \label{fig:examples}
\end{figure}

A fungal automaton is a one-dimensional cellular automaton with binary cell states and ternary, including central cell, cell neighbourhood, governed by two elementary cellular automata (ECA) rules, namely the cell state transition rule $f$ and the Woronin bodies adjustment rule $g$:
$\mathcal{M}=\langle \mathbf{N}, u, \mathbf{Q}, f, g \rangle$. Each cell $x_i$ has a unique index $i \in \mathbf{N}$. Its state is updated from $\mathbf{Q}=\{0, 1\}$ in discrete time depending of its current state $x^t_i$, the states of its left $x^t_{i-1}$ and right neighbours $x^t_{i+1}$ and the state of cell $x$'s Woronin body $w$. Woronin bodies take states from $\mathbf{Q}$: $w^t=1$ means Woronin bodies (Fig.~\ref{fig:bioscheme}) in cell $x$ blocks the pores and the cell has no communication with its neighbours, and $w^t=0$ means that Woronin bodies in cell $x$ do not block the pores. Woronin bodies update their states $g(\cdot)$, $w^{t+1}=g(u(x)^t)$, depending on the state of neighbourhood $u(x)^t$. 
Cells $x$ update their states by function $f(\cdot)$ if their Woronin bodies do not block the pores. 

Two species of mycelium automata are considered $\mathcal{M}_1$, where each cell updates its state as following:
$$
x^{t+1}=
\begin{cases} 
0 &\mbox{if } w^t=1 \\ 
f(u(x)^t) & \mbox{otherwise} 
\end{cases} 
$$
and $\mathcal{M}_2$ where each cell updates its state as following:
$$
x^{t+1}=
\begin{cases} 
x^t &\mbox{if } w^t=1 \\ 
f(u(x)^t) & \mbox{otherwise} 
\end{cases} 
$$
where $w^{t}=g(u(x)^t)$.

State `1' in the cells of array $x$ symbolises metabolites, signals exchanged between cells. Where pores in a cell are open the cell updates its state by ECA rule $f: \{0,1\}^3 \rightarrow \{0,1\}$. 

In automaton $\mathcal{M}_1$, when Woronin bodies block the pores in a cell, the cell does not update its state and remains in the state `0' and left and right neighbours of the cells can not detect any `cargo' in this cell. In automaton, $\mathcal{M}_2$, where Woronin bodies block the pores in a cell, the cell does not update its state and remains in its current state. In real living mycelium glucose and possibly other metabolites~\cite{bleichrodt2015switching} can still cross the septum even when septa are closed by Woronin bodies, but we can ignore this fact in present abstract model. 

Both species are biologically plausible and, thus, will be studied in parallel. The rules for closing and opening Woronin bodies are also ECA rules $g: \{0,1\}^3 \rightarrow \{0,1\}$. If $g(u(x)^t)=0$ this means that pores are open, if $g(u(x)^t)=1$ Woronin bodies block the pores. 
Examples of space-time configurations of both species of $\mathcal{M}$ are shown in Fig.~\ref{fig:examples}.

\section{Properties of composition $f \circ g$}
\label{composition}

\subsubsection*{Predecessor sets}

\begin{figure}[!tbp]
\centering
 \subfigure[]{\includegraphics[scale=0.85]{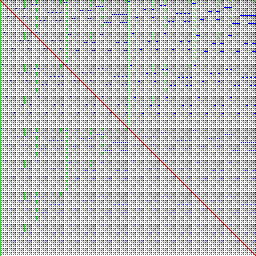}}
  \subfigure[]{\includegraphics[scale=0.85]{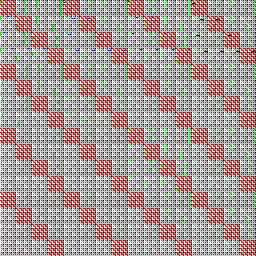}}
    \subfigure[]{\includegraphics[scale=0.68]{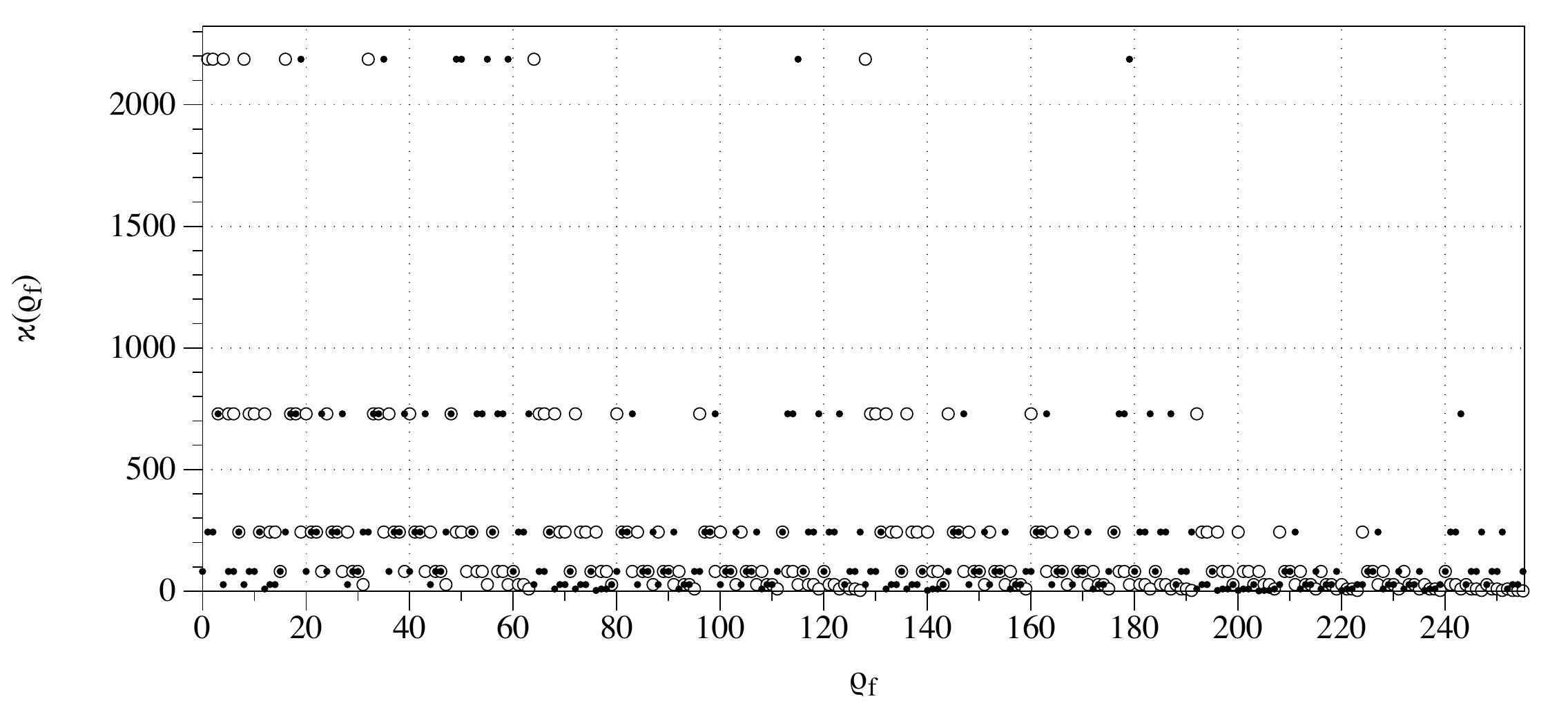}}
    \caption{Mapping $\mathbf{F} \times \mathbf{F} \rightarrow \mathbf{F}$ for automaton $\mathcal{M}_1$~(a) and $\mathcal{M}_2$~(b) is visualised as an array of pixels, $\mathbf{P}=(p)_{0 \leq \rho_f \leq 255, 0 \leq \rho_f \leq 255}$. An entry at the intersection of any  $\rho_f$ and $\rho_g$ is a coloured as follows: red if $p_{\rho_f \rho_g}=p_{\rho_g \rho_f}$, blue if $\rho_g=p_{\rho_g \rho_f}$, green if $\rho_f=p_{\rho_g \rho_f}$. 
    (c)~Sizes of $\mathbf{P}(h)$  sets for $\mathcal{M}_1$, circle, and $\mathcal{M}_2$, solid discs, are shown for every function $h$ apart of rule 0 ($\mathcal{M}_1$) and rule 51 ($\mathcal{M}_2$). }
    \label{fig:sizes}
\end{figure}

 Let $\mathbf{F}=\{h: \{0,1\}^3 \rightarrow \{0,1\} \}$ be a set of all ECA functions. Then for any composition $f\circ g$, where $f, g \in \mathbf{F}$, can be converted to a single function $h \in \mathbf{F}$. For each $h \in \mathbf{F}$ we can construct a set $\mathbf{P}(h)=\{f\circ g \in \mathbf{F} \times \mathbf{F}\,|\, f\circ g \rightarrow h \}$.
The sets $\mathbf{P}(h)$ for each $h \in \mathbf{F}$ are available online\footnote{\url{https://figshare.com/s/b7750ee3fe6df7cbe228}}. 

\begin{table}[!tbp]
\centering
    \subfigure[Rules per  $|\mathbf{P}(h)|$]{
    \begin{tabular}{l|l}
$\sigma$ & $\gamma$ \\ \hline
1 & 1\\
3 & 8\\
9 & 28\\
27 & 56\\
81 & 70\\
243 & 56\\
729 & 28\\
2187 & 8\\
6561 & 1
\end{tabular}} \hspace{10mm}
\subfigure[$\mathcal{M}_1$: Class III rules]{
\begin{tabular}{p{2cm}|l}
Rule & $\sigma$ \\ \hline
18 & 729 \\
22, 146 & 243\\
30, 45, 60, 90, 105, 150	& 81\\
122	& 27\\
126	& 9\\
\end{tabular}
}
\subfigure[$\mathcal{M}_1$: Class IV rules]{
\begin{tabular}{p{2cm}|l}
Rule & $\sigma$ \\ \hline
41 & 243\\
54, 106, 110 & 81\\
\end{tabular}
}\\
\subfigure[$\mathcal{M}_2$: Class III rules]{
\begin{tabular}{p{2cm}|l}
Rule & $\sigma$ \\ \hline
18 & 729\\
22, 146 & 243\\
30, 45, 60 90, 105, 150 & 81\\
122 & 243\\
126 & 81
\end{tabular}
}
\subfigure[$\mathcal{M}_2$: Class IV rules]{
\begin{tabular}{p{2cm}|l}
Rule & $\sigma$ \\ \hline
41 & 243\\
54 & 729\\
106 & 81\\
110 & 27
\end{tabular}
}
    \caption{Characterisations of automaton mapping $\mathbf{F} \times \mathbf{F} \rightarrow \mathbf{F}$. (a)~Size $\sigma$ of $\mathbf{P}(h)$ vs a number $\gamma$ of functions $h$ having set $\mathbf{P}(h)$ of size $\sigma$. T (b)~Sizes of sets $\mathbf{P}(h)$ for rules from Wolfram class III. (b)~Sizes of sets $\mathbf{P}(h)$ for rules from Wolfram class IV. }
    \label{tab:sizes}
\end{table}

A size of $\mathbf{P}(h)$ for each $h$ is shown in Fig.~\ref{fig:sizes}c. The functions with largest size of $\mathbf{P}(h)$ are rule  0 in automaton $\mathcal{M}_1$  and rule 51 (only neighbourhood configurations (010, 011, 110, 111 are mapped into 1) in $\mathcal{M}_2$.  

Size $\sigma$ of $\mathbf{P}(h)$ vs a number $\gamma$ of functions $h$ having set $\mathbf{P}(h)$ of size $\sigma$ is shown for automata $\mathcal{M}_1$ and $\mathcal{M}_2$ in Table~\ref{tab:sizes}a.

With regards to Wolfram classification~\cite{wolfram1994cellular}, sizes of $\mathbf{P}(h)$ for rules from Class III vary from 9 to 729 in $\mathcal{M}_1$ (Tab.~\ref{tab:sizes}b). Rule 126 would be the most difficult to obtain in $\mathcal{M}_1$ by composition two ECA rules chosen at random, it has only 9 `predecessor' $f\circ g$ pairs. Rule 18 would be the easiest, for Class III rules, to be obtained, it has 729 predecessors, in both $\mathcal{M}_1$ (Tab.~\ref{tab:sizes}b) and $\mathcal{M}_2$ (Tab.~\ref{tab:sizes}d). In $\mathcal{M}_1$, one rule, rule 41, from the class IV has 243 $f\circ g$ predecessors, and all other rules in that class have 81 (Tab.~\ref{tab:sizes}c). From Class IV rule 54 has the largest number of predecessors in $\mathcal{M}_2$, and thus can be considered as most common (Tab.~\ref{tab:sizes}d).

\begin{table}[!tbp]
    \centering
    \begin{tabular}{c|p{12cm}}
$f\circ f$     & $f$ \\ \hline
0 & 0, 1, 2, 3, 16, 17, 18, 19, 32, 33, 34, 35, 48, 49, 50, 51\\
1 & 128, 129, 130, 131, 144, 145, 146, 147, 160, 161, 162, 163, 176, 177, 178, 179\\
2 & 64, 65, 66, 67, 80, 81, 82, 83, 96, 97, 98, 99, 112, 113, 114, 115\\
3 & 192, 193, 194, 195, 208, 209, 210, 211, 224, 225, 226, 227, 240, 241, 242, 243\\
16 & 8, 9, 10, 11, 24, 25, 26, 27, 40, 41, 42, 43, 56, 57, 58, 59\\
17 & 136, 137, 138, 139, 152, 153, 154, 155, 168, 169, 170, 171, 184, 185, 186, 187\\
18 & 72, 73, 74, 75, 88, 89, 90, 91, 104, 105, 106, 107, 120, 121, 122, 123\\
19 & 200, 201, 202, 203, 216, 217, 218, 219, 232, 233, 234, 235, 248, 249, 250, 251\\
32 & 4, 5, 6, 7, 20, 21, 22, 23, 36, 37, 38, 39, 52, 53, 54, 55\\
33 & 132, 133, 134, 135, 148, 149, 150, 151, 164, 165, 166, 167, 180, 181, 182, 183\\
34 & 68, 69, 70, 71, 84, 85, 86, 87, 100, 101, 102, 103, 116, 117, 118, 119\\
35 & 196, 197, 198, 199, 212, 213, 214, 215, 228, 229, 230, 231, 244, 245, 246, 247\\
48 & 12, 13, 14, 15, 28, 29, 30, 31, 44, 45, 46, 47, 60, 61, 62, 63\\
49 & 140, 141, 142, 143, 156, 157, 158, 159, 172, 173, 174, 175, 188, 189, 190, 191\\
50 & 76, 77, 78, 79, 92, 93, 94, 95, 108, 109, 110, 111, 124, 125, 126, 127\\
51 & 204, 205, 206, 207, 220, 221, 222, 223, 236, 237, 238, 239, 252, 253, 254, 255\\
    \end{tabular}
    \caption{Diagonals of automaton $\mathcal{M}_2$.}
    \label{tab:diagonalsM2}
\end{table}

\subsubsection*{Diagonals} In automaton $\mathcal{M}_1$ for any  $f \in \mathbf{F}$ $f \circ f = 0$.   Assume $f: \{0,1\}^3 \rightarrow 1$ then Woronin bodies close the pores and, thus, second application of $f$ produces state `0'. If $f: \{0,1\}^3 \rightarrow 0$ then Woronin bodes does not close pores but yet second application of the $f$ produce state `0'.

For automaton $\mathcal{M}_2$ a structure of diagonal mapping $f \circ f \rightarrow h$, where $f, h \in \mathbf{F}$ is shown in Tab.~\ref{tab:diagonalsM2}. The set of the diagonal outputs $f \circ f$ consists of 16 rules:
(0, 1, 2, 3), (16, 17, 18, 19), (32, 33, 34, 35), (48, 49, 40, 51). These set of rules can be reduced to the following rule. Let $C(x^t)=[u(x)^t=(111)] \vee [u(x)^t=(111)]$ and $B(x^t)=[u(x)^t=(011)] \vee [u(x)^t=(010)]$. Then 
$x^t=1$ if $C(x)^t \vee C(x)^t \wedge B(x^t)$. 

\subsubsection*{Commutativity} 

In automaton $\mathcal{M}_1$, for any  $f, g \in \mathbf{F}$ $f \circ g \neq g \circ f$ only if $f \neq g$. In automaton  $\mathcal{M}_2$ there are 32768 pairs of function which $\circ$ is commutative, their distribution visualised in red in  Fig.~\ref{fig:sizes}b.

\subsubsection*{Identities and zeros} 

In automaton $\mathcal{M}_1$ there are  no left or right identities, neither right zeros in $\langle \mathbf{F}, \mathbf{F}, \circ \rangle$. The only left zero is the rule 0. In automaton $\mathcal{M}_2$ there are no identities or zeros at all.

\subsubsection*{Associativity} 

In automaton $\mathcal{M}_1$ there 456976 triples $\langle f, g, h \rangle$ on which operation $\circ$ is associative: $(f \circ g) \circ h = f \circ (g \circ h)$. The ratio of associative triples to the total number of triples is 0.027237892. 
There are 104976 associative triples in $\mathcal{M}_2$, a ratio of 0.006257057.

\section{Tuning Complexity: Rule 110}
\label{complexity}

\begin{figure}[!tbp]
    \centering
    \subfigure[]{\includegraphics[scale=0.48]{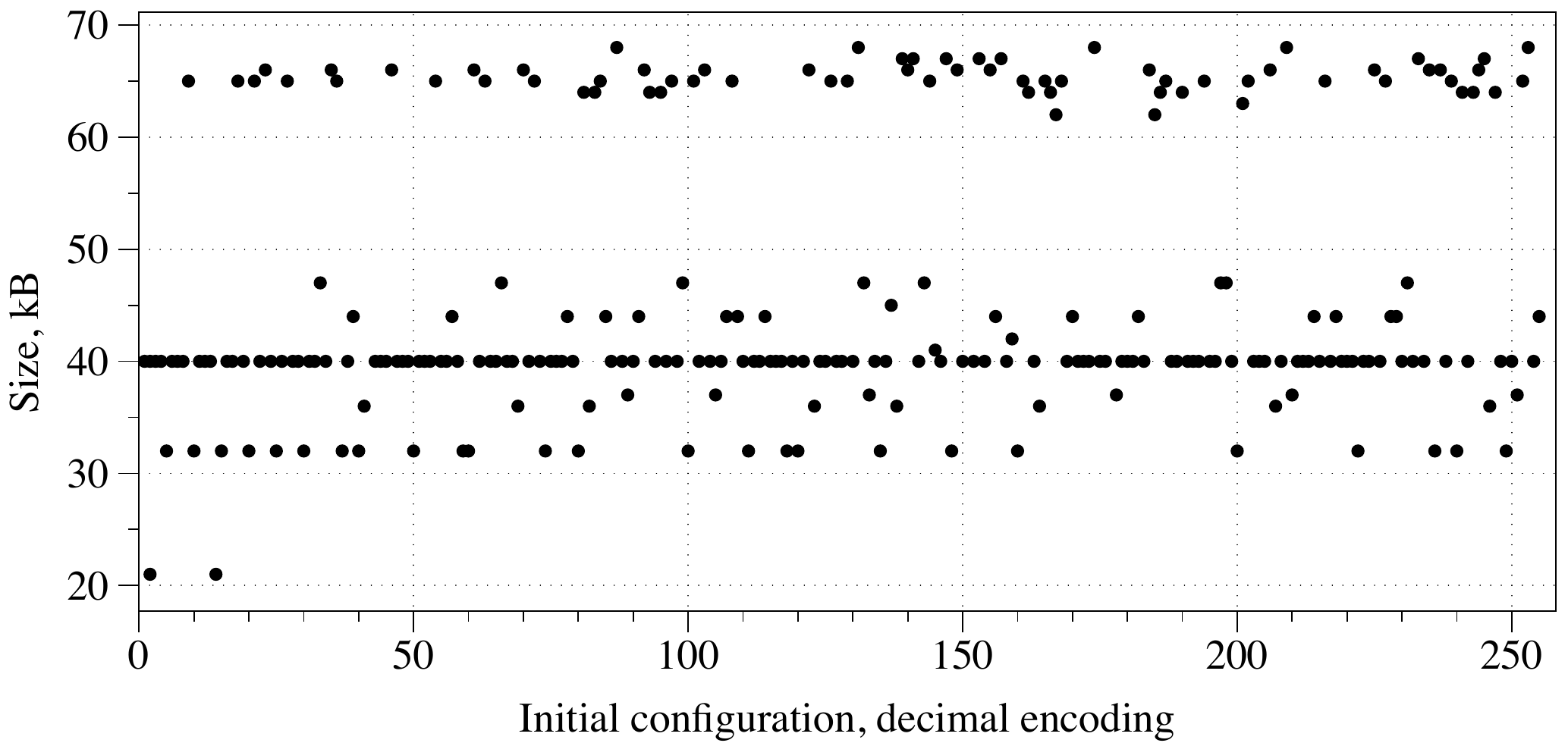}}
    \subfigure[]{\includegraphics[width=0.38\textwidth]{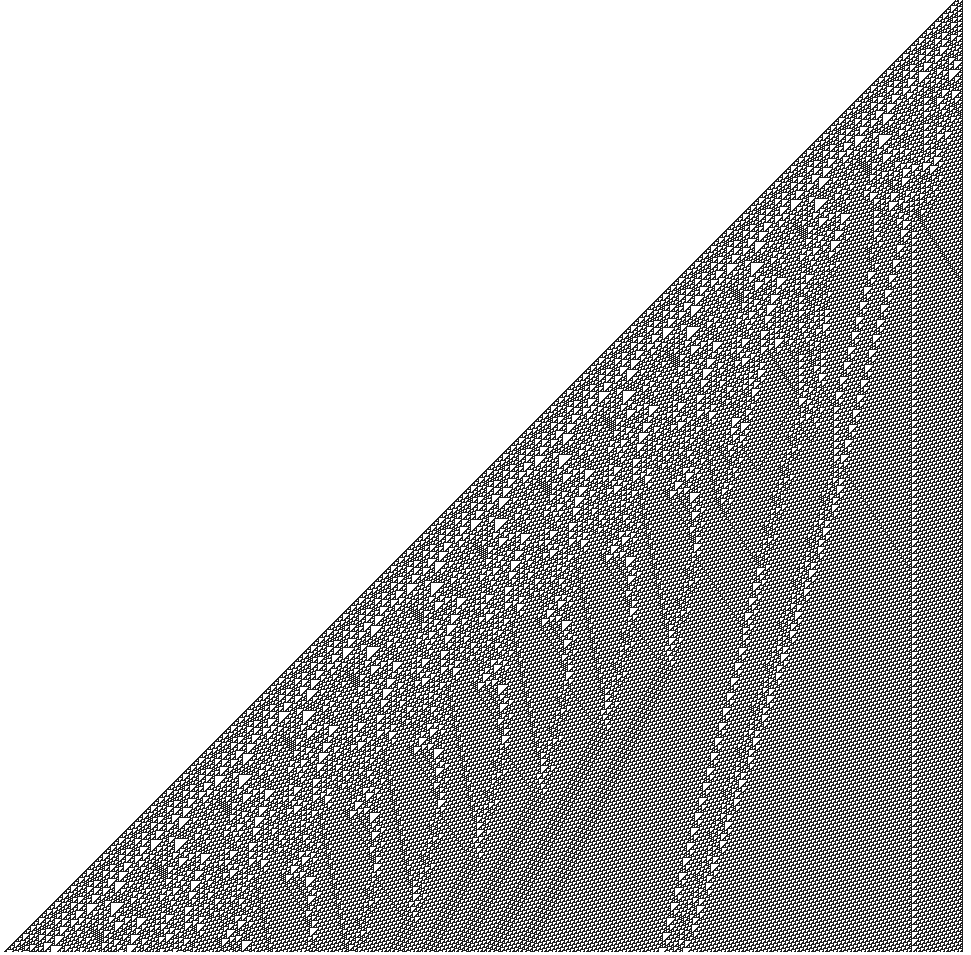}}
     \subfigure[]{\includegraphics[width=0.38\textwidth]{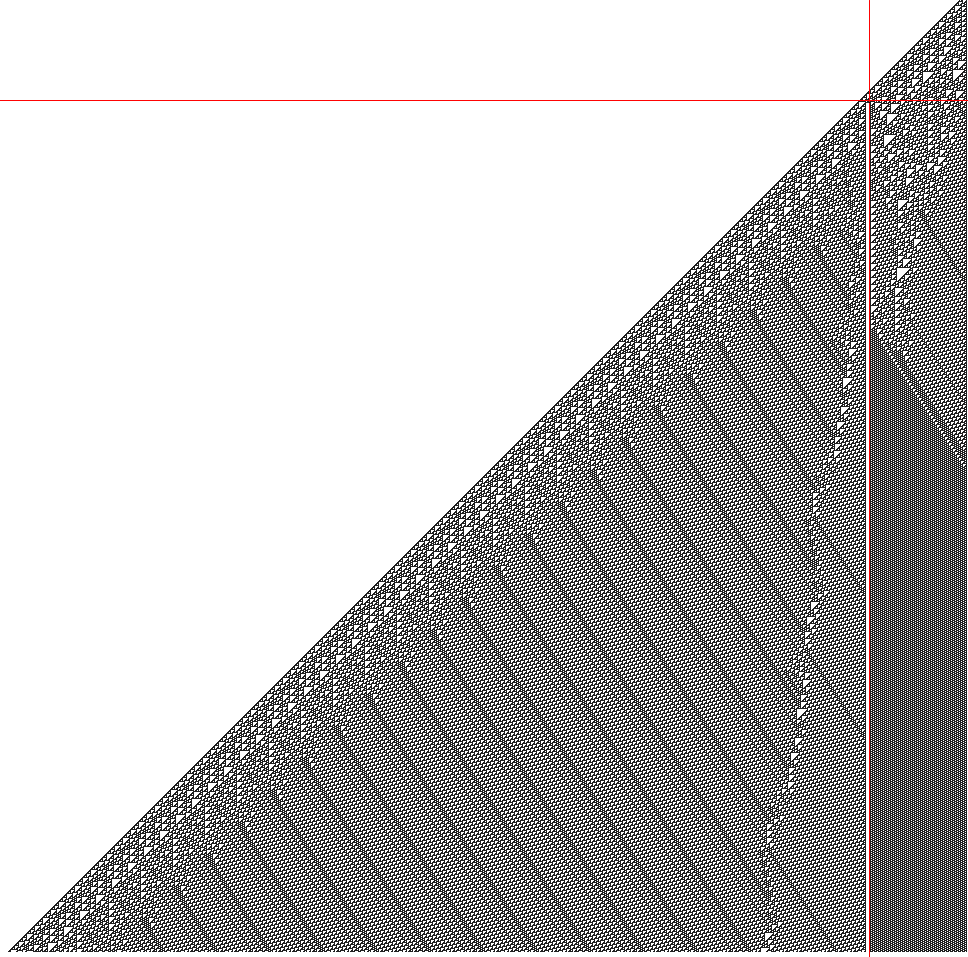}}
\subfigure[]{\includegraphics[width=0.49\textwidth]{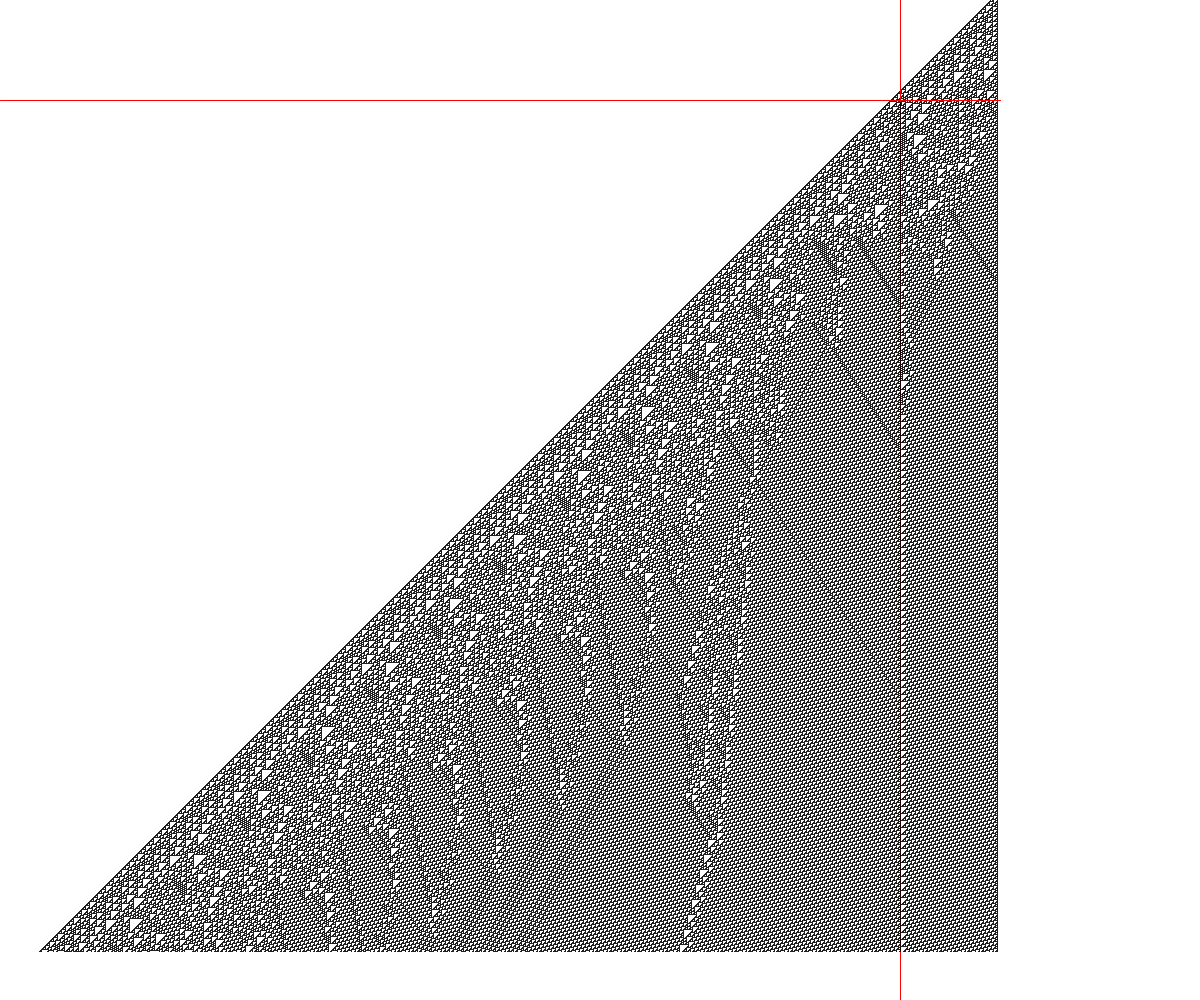}}
    \caption{(a)~Estimates of LZ complexity of space-time configurations of ECA Rule 110 without Woronin bodies. (b)~A space-time configuration of ECA Rule 110 evolving from initial configuration 10110001 (177), no Woronin bodies are activated. (c)~A space-time configuration of $\mathcal{M}_1$ Rule 110 evolving from initial configuration 10110001 (177), Woronin body is governed by rule 43; red lines indicate time when the body was activated and position of the cell with the body. In (bcd), a pixel in position $(i,t)$ is black if $x_i^t=1$.}
    \label{fig:LZSingleConf}
\end{figure}

\begin{figure}[!tbp]
    \centering
    \includegraphics[scale=0.25]{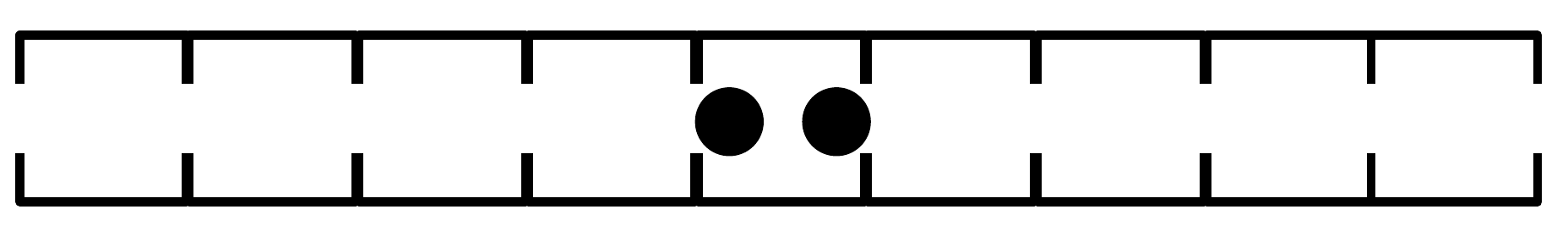}
    \caption{Only one cell has Woronin body.}
    \label{fig:oneWoronin}
\end{figure}

To evaluate on how introduction of Woronin bodies could affect complexity of automaton evolution, we undertook two series of experiments. In the first series we used fungal automaton where just one cell has a Woronin body (Fig.~\ref{fig:oneWoronin}). In the second series we employed fungal automaton where regularly positioned cells (but not all cells of the array) have Woronin bodies. 

State transition functions $g$ of Woronin bodies were varied across the whole diapason but the state transition function $f$ of a cell was Rule 110, $\rho_f=110$. We have chosen Rule 110 because the rule is proven to be computationally universal~\cite{lindgren1990universal,cook2004universality}, P-complete~\cite{neary2006p}, the rules belong to Wolfram class IV renown for exhibiting complex and non-trivial interactions between travelling localisation~\cite{wolfram1984universality}, rich families of gliders can be produce in collisions with other gliders~\cite{martinez2003production, martinez2006gliders,mora2007rule}.

We wanted to check how an introduction of Woronin bodies affect dynamics of most complex space-time developed of Rule 110 automaton. Thus, we evolved the automata from all possible initial configurations of 8 cells placed near the end of $n=1000$ cells array of resting cells and allowing to evolve for 950 iterations. Lempel--Ziv complexity (compressibility) LZ was evaluated via sizes of  space-time configurations saved as PNG files. This is sufficient because the 'deflation' algorithm used in  PNG lossless compression~\cite{roelofs1999png,howard1993design, deutsch1996zlib}  is a variation of the classical Lempel--Ziv 1977 algorithm~\cite{ziv1977universal}.
Estimates of LZ complexity for each of 8-cell initial configurations are shown in Fig.~\ref{fig:LZSingleConf}a. The initial configurations with highest estimated LZ complexity are 10110001 (decimal 177), 11010001 (209), 10000011 (131), 11111011 (253), see example of space-time dynamics in Fig.~\ref{fig:LZSingleConf}b. 

We assumed that a cell in the position $n-100$ has a Woronin body which can be activated (Fig.~\ref{fig:oneWoronin}), i.e. start updating its state by rule $f$, after 100th iteration of the automaton evolution. We then run 950 iteration of automaton evolution for each of 256 Woronin rules and estimated LZ complexity. In experiments with $\mathcal{M}_1$ we found that 128 rules, with even decimal representations, do not affect space time dynamics of evolution and 128 rules, with even decimal representations, reduce complexity of the space-time configuration. The key reasons for the complexity reduction (compare Fig.~\ref{fig:LZSingleConf}b and c) are cancellation of three gliders at c. 300th iteration and simplification of the behaviour of glider guns positioned at the tail of the propagating wave-front. In experiments with $\mathcal{M}_2$ 128 rules, with even decimal representations, do not change the space-time configuration of the author. Other 128 rules reduce complexity and modify space-time configuration by re-arranging the structures of glider guns and establishing one oscillators at the site surrounding position of the cell with Woronin body (Fig.~\ref{fig:LZSingleConf}d).

\begin{figure}[!tbp]
    \centering
    \subfigure[]{\includegraphics[width=0.2\textwidth]{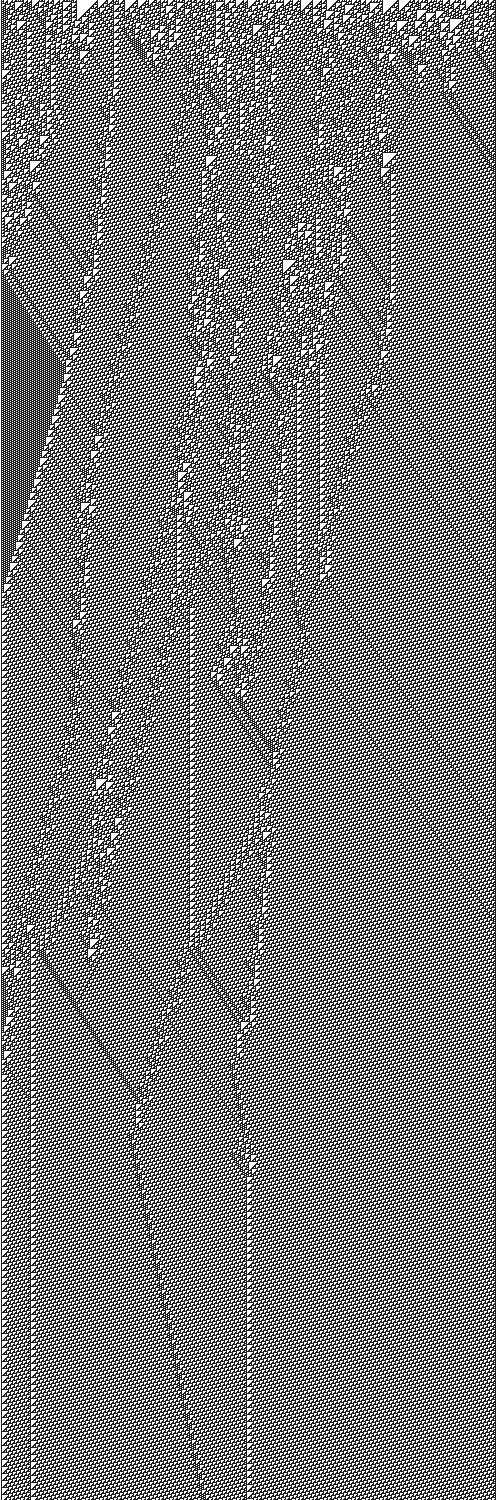}}
    \subfigure[$\mathcal{M}_1$, $\rho_g=133$]{\includegraphics[width=0.2\textwidth]{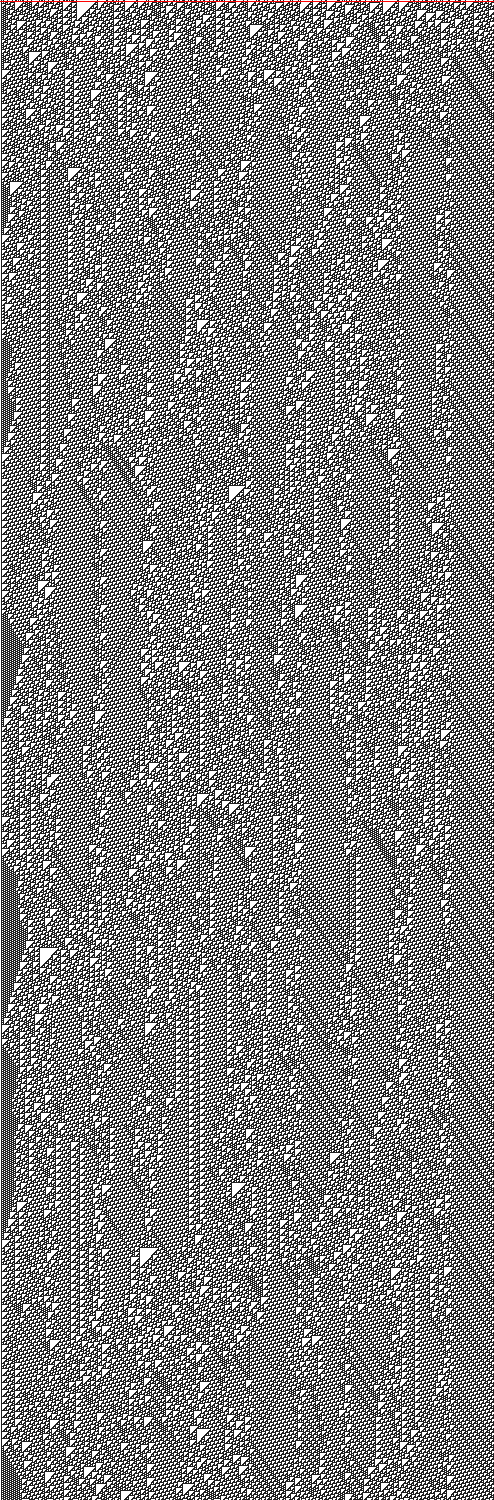}}
\subfigure[$\mathcal{M}_1$, $\rho_g=29$]{\includegraphics[width=0.2\textwidth]{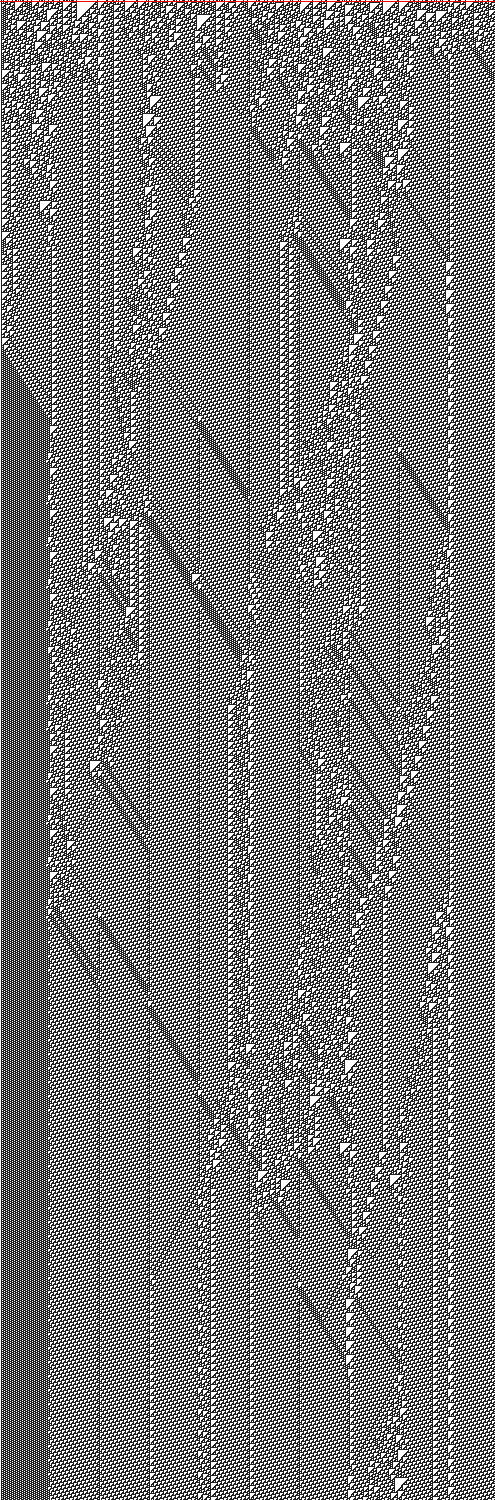}} 
    \subfigure[$\mathcal{M}_1$, $\rho_g=49$]{\includegraphics[width=0.2\textwidth]{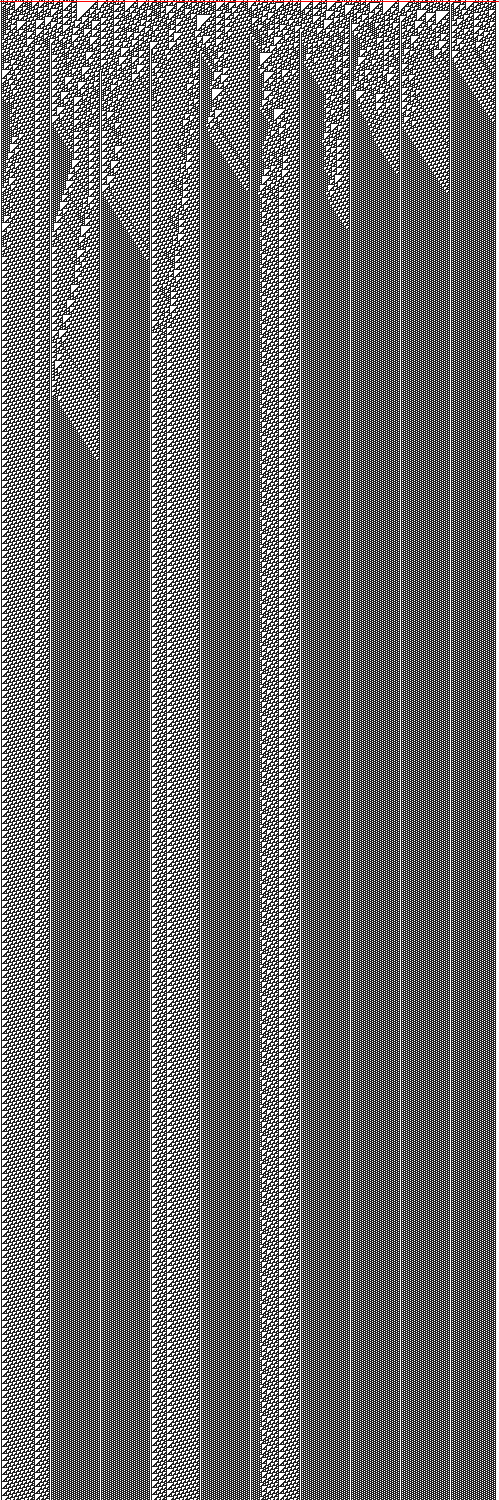}}
\subfigure[$\mathcal{M}_2$, $\rho_g=193$]{\includegraphics[width=0.2\textwidth]{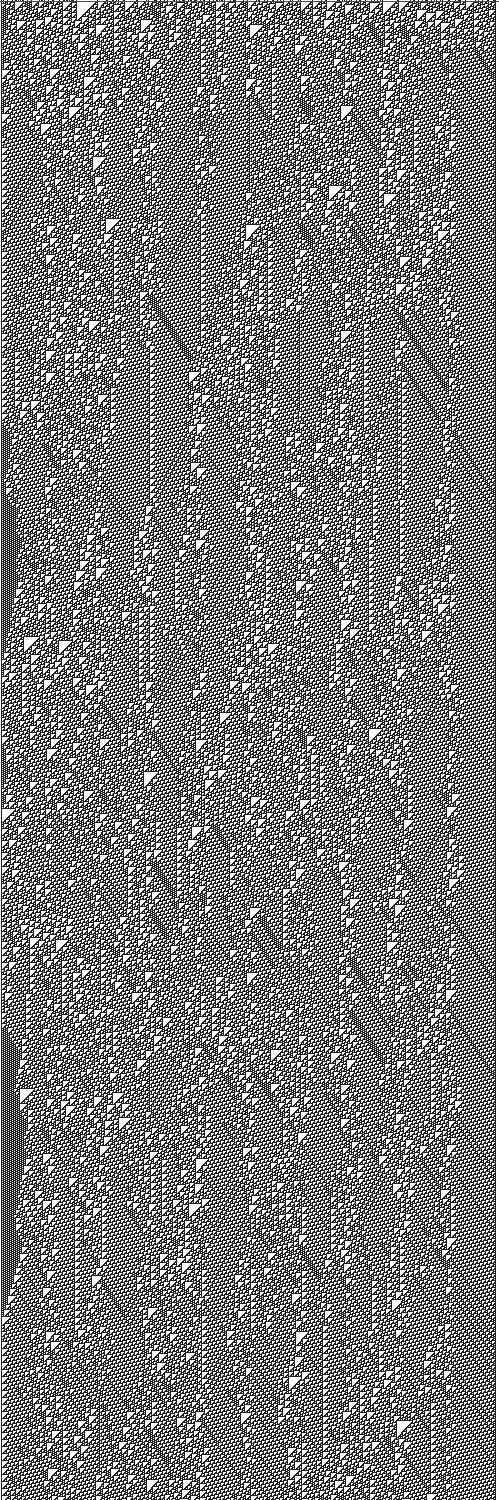}}
\subfigure[$\mathcal{M}_2$, $\rho_g=5$]{\includegraphics[width=0.2\textwidth]{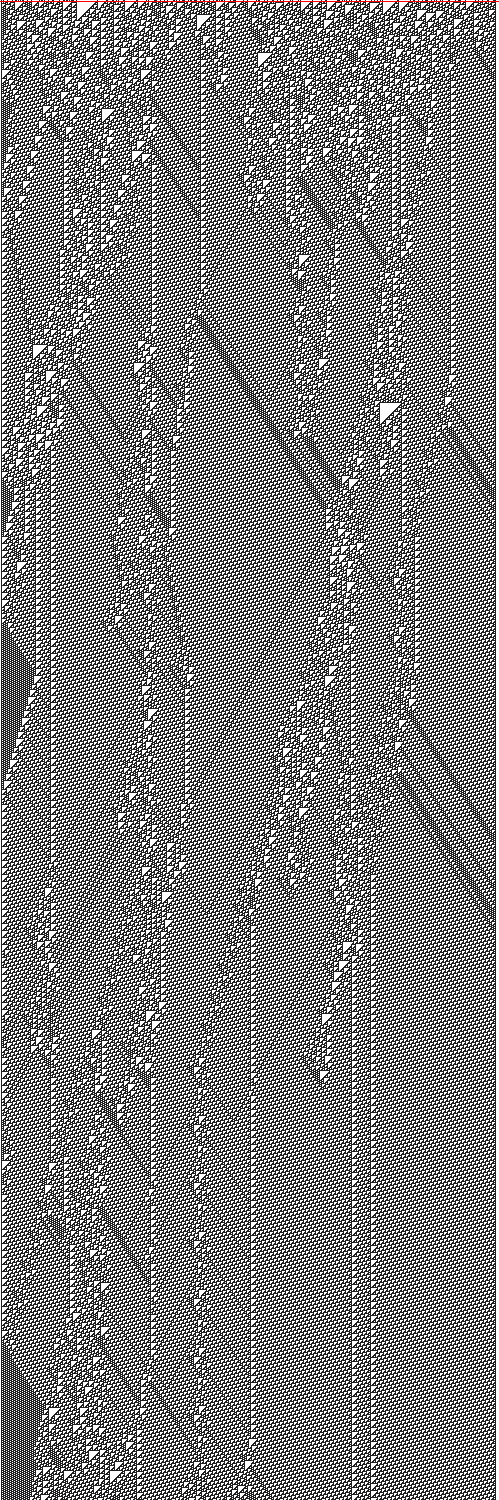}}
\subfigure[$\mathcal{M}_2$, $\rho_g=221$]{\includegraphics[width=0.2\textwidth]{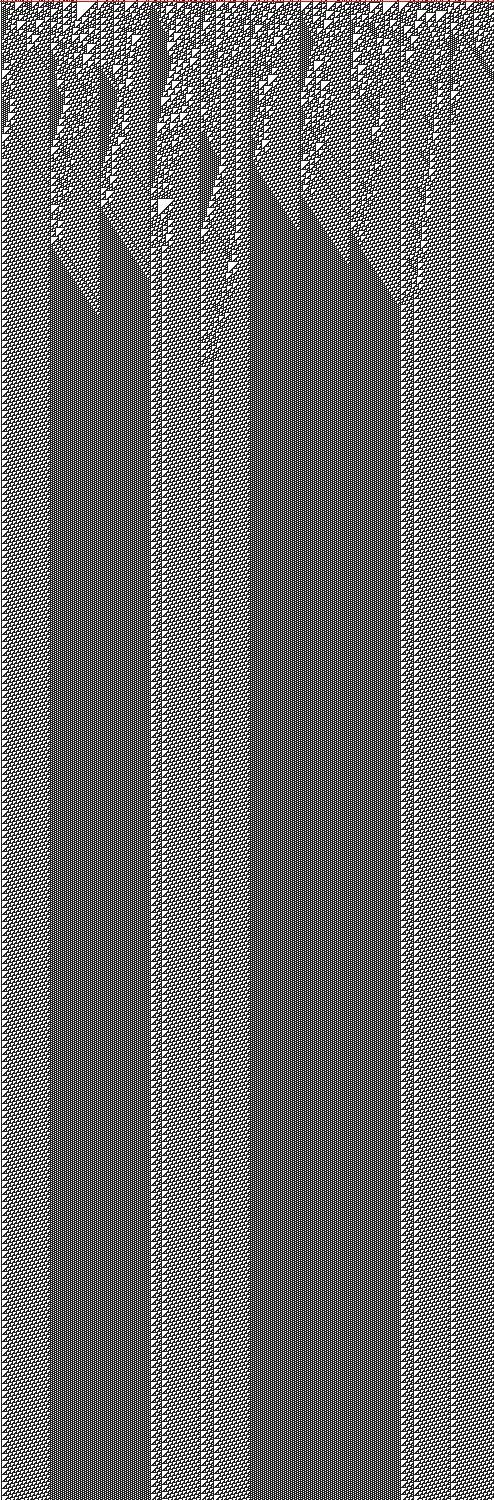}}
\subfigure[$\mathcal{M}_2$, $\rho_g=174$]{\includegraphics[width=0.2\textwidth]{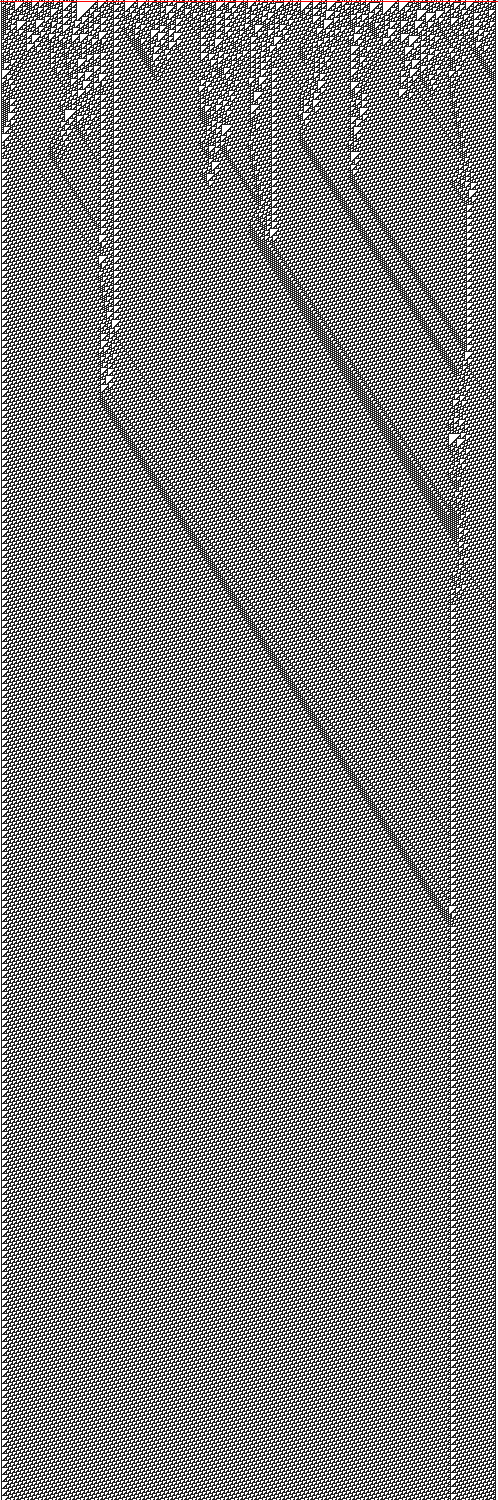}}
    \caption{(a)~ECA Rule 110, no Woronin bodies. Space-time evolution of $\mathcal{M_1}$~(bcd) and $\mathcal{M_2}$~(e--h) for Woronin rules shown in subcaption. LZ complexity of space-time configurations decreases from (b) to (d) and from (e) to (h). Every 50th cell has a Woronin body.} 
    \label{fig:CArule110}
\end{figure}

\begin{figure}[!tbp]
    \centering
    \includegraphics[scale=0.6]{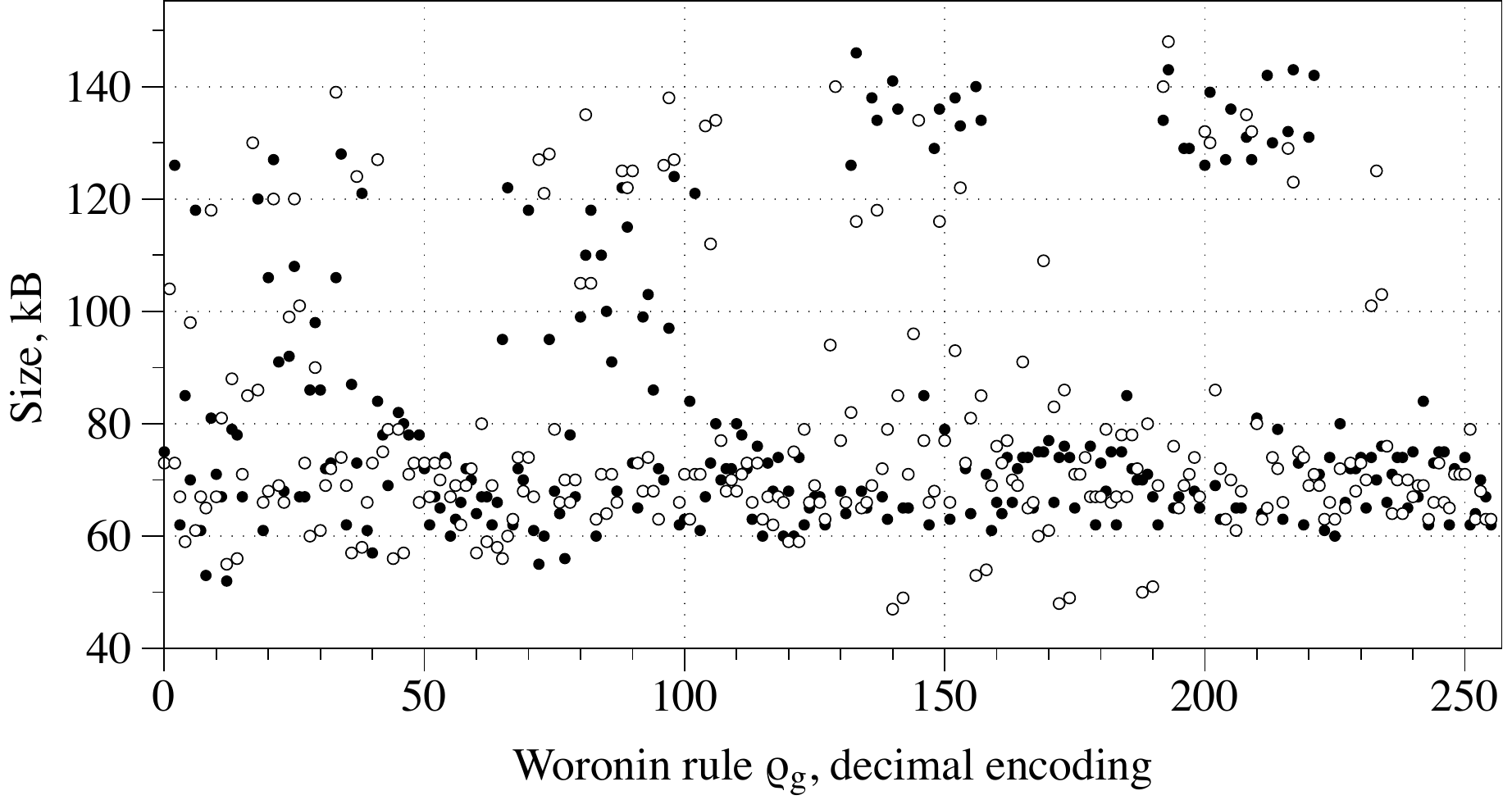}
    \caption{Estimations of LZ complexity of space-time, 500 cells by 500 iterations, configurations of $\mathcal{M}_1$, discs, and $\mathcal{M}_1$, circles, for all Woronin functions $g$.}
    \label{fig:intervalcompplexity}
\end{figure}

In second series of experiments we regularly positioned cells with Woronin bodies along the 1D array: every 50th cell has a Woronin body. Then we evolved fungal automata $\mathcal{M}_1$ and $\mathcal{M}_2$ from exactly  the same initial random configuration with density of `1' equal to 0.3. Space-time configuration of the automaton without Woronin bodies is shown in Fig.~\ref{fig:CArule110}a. Exemplar of space-time configurations of automata with Woronin bodies are shown in Fig.~\ref{fig:CArule110}b--h.   As seen in Fig.~\ref{fig:intervalcompplexity} both species of fungal automata show similar dynamics of complexity along the Woronin transition functions ordered by their decimal values. The automaton $\mathcal{M}_1$ has average LZ complexity 82.2 ($\overline{\sigma}=24.6$) and the automaton $\mathcal{M}_2$ 78.4 ($\overline{\sigma}=22.1$).
Woronin rules $g$ which generate most LZ complex space-time configurations are $\rho_g=133$ in $\mathcal{M})_1$ (Fig.~\ref{fig:CArule110}b)  and $\rho_g=193$ in $\mathcal{M})_2$ (Fig.~\ref{fig:CArule110}e). The space-time dynamics of the automaton is characterised by a substantial number of gliders guns and gliders (Fig.~\ref{fig:CArule110}b).  Functions being in the middle of the descending hierarchy of LZ complexity produce space-time configurations with declined number of travelling localisation and growing domains of homogeneous states (Fig.~\ref{fig:CArule110}cg). Automata with Woronin functions at the bottom of the complexity hierarchy quickly (i.e. after 200-300 iterations) evolve towards stable, equilibrium states (Fig.~\ref{fig:CArule110}dh).

\section{Local events}
\label{localevents}

\begin{figure}[!tbp]
    \centering
    \subfigure[$\mathcal{M}_1$, $\rho_g=2$]{\includegraphics[scale=0.4]{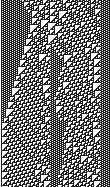}}
    \subfigure[$\mathcal{M}_1$, $\rho_g=15$]{\includegraphics[scale=0.4]{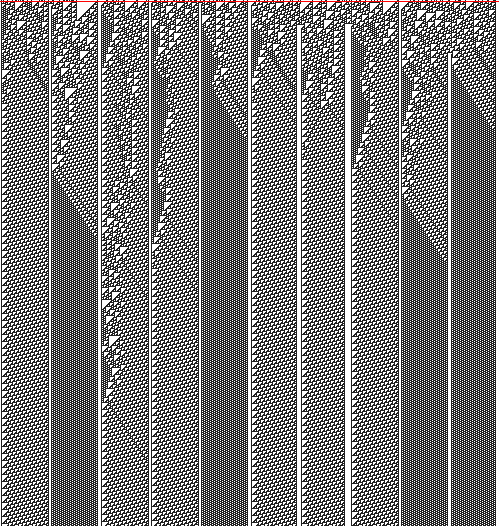}}
       \subfigure[$\mathcal{M}_1$, $\rho_g=21$]{\includegraphics[scale=0.4]{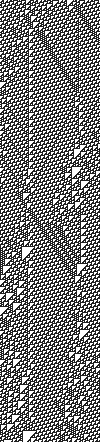}}
         \subfigure[$\mathcal{M}_1$,
    $\rho_g=31$]{\includegraphics[scale=0.4]{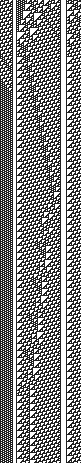}}
    \subfigure[$\mathcal{M}_1$,
    $\rho_g=29$]{\includegraphics[scale=0.4]{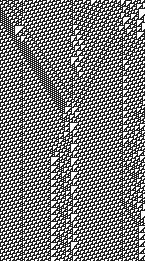}}
    \subfigure[$\mathcal{M}_1$,
    $\rho_g=201$]{\includegraphics[scale=0.4]{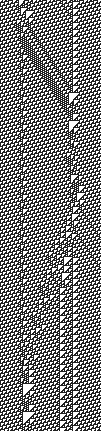}}
    \caption{
    (a)~Localisation travelling left was stopped by the Woronin body. 
    (b)~Analog of a memory register. 
    (c)~Reflections of travelling localisations from cells with Woronin bodies. 
    (d)~Modification of glider state in the vicitinity of Woronin bodies.
    (e)~A fragment of configuration of automaton with $\rho_g=29$, left cell states, right Woronin bodies states. 
    (f)~Enlarged sub-fragment of the fragment (d) where Wonorin body tunes the outcome of the collision.
    For both automata $\rho_f=110$.}
    \label{fig:scenarios}
\end{figure}

Let us consider some local events happening in the fungal automata discussed in Sect.~\ref{complexity}: every 50th cell of an array has a Woronin body.

\textbf{Retaining gliders.} A glider can be stopped and converted into a station localisation by a cell with Woronin body. As exemplified in Fig.~\ref{fig:scenarios}a, the localisation travelling left was stopped from further propagation by a cell with Woronin body yet the localisation did not annihilate but remained stationary. 

\textbf{Register memory.} Different substrings of input string (initial configuration) might lead to different equilibrium configurations achieved in the domains of the array separated by cells with Woronin bodies.  When there is just two types of equilibrium configurations they be seen as `bit up' and `bit down' and therefore such fungal automaton can be used a memory register (Fig.~\ref{fig:scenarios}b). 

\textbf{Reflectors.} In many cases cells with Woronin bodies induce local domains of stationary, sometimes time oscillations, inhomogeneities which might act as reflectors for travelling localisations. An example is shown in Fig.~\ref{fig:scenarios}c where several localisations are repeatedly bouncing between two cells with Woronin bodies.

\textbf{Modifiers.} Cells with Woronin bodies can act as modifiers of states of gliders reflected from them and 
of outcomes of collision between travelling localizations.
In Fig.~\ref{fig:scenarios}d we can see how a travelling localisation is reflected from the vicinity of Woronin bodies three times: every time the state of the localisation changes. On the third reflection the localisation becomes stationary. 
In the fragment (Fig.~\ref{fig:scenarios}e) of space-time configuration of automaton with Woronin bodies governed by $\rho_g=201$ of the fragment we can see how two localisations got into contact with each in the vicinity of the Woronin body and an advanced structure is formed two breathing stationary localisations act as mirror, and there are streams of travelling localisations between them. A multi-step chain reaction can be observed in 
Fig.~\ref{fig:scenarios}f: there are two stationary, breathing, localisations at the sites of the cells with Woronin bodies. A glider is formed on the left stationary localisation. The glider travel to the right and collide into right breather. In the result of the collision the breath undergoes structural transitions, emits a glider travelling left and transforms itself into a pair of stationary breathers. Meantime the newly born glider collided into left breather and changes its state.

\section{Discussion}
\label{conclusion}

\begin{figure}[!tbp]
    \centering
    \includegraphics[scale=0.3]{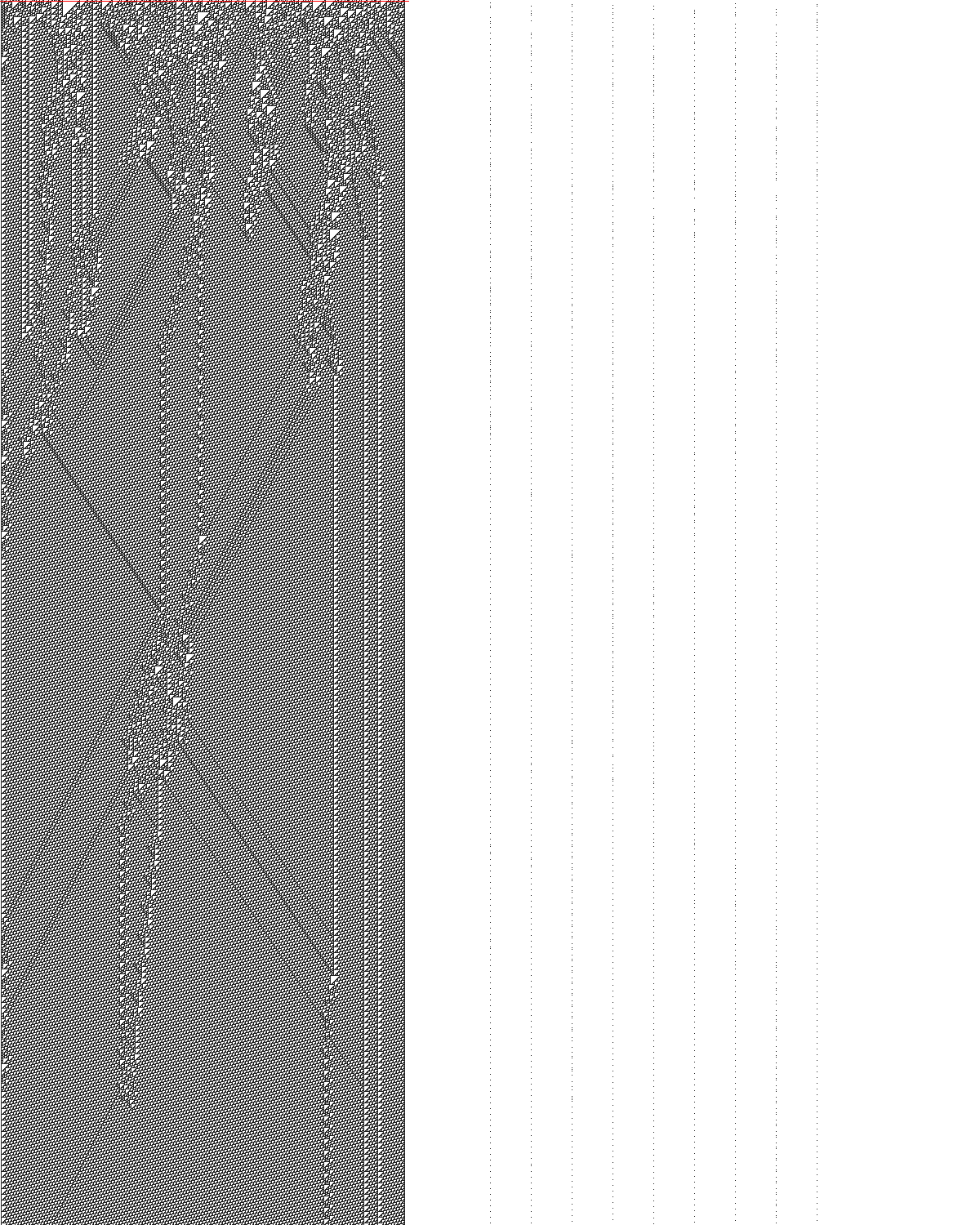}
    \caption{An example of 5-inputs-7-outputs collision in $\mathcal{M}_2$, $\rho_f=110$, $\rho_g=40$. Every 50th cell has a Woronin body. Cells state transitions are shown on the left, Woronin bodies state transitions on the right. A pixel in position $(i,t)$ is black if $x_i^t=1$, left, or $w_i^t=1$, right.}
    \label{fig:complexcollision}
\end{figure}

As a first step towards formalisation of fungal intelligence we introduced one-dimensional fungal automata operated by two local transition function: one, $g$, governs states of Woronin bodies (pores are open or closed), another, $f$, governs cells states: `0' and `1'. We provided a detailed analysis of the magma $\langle f, g, \circ \rangle$, results of which might be useful for future designs of computational and language recognition structures with fungal automata. The magma as a whole does not satisfy any other property but closure. Chances are high that there are subsets of the magma which might satisfy conditions of other algebraic structures. A search for such subsets could be one of the topics of further studies. 

Another topic could be an implementation of computational circuits in fungal automata. For certain combination of $f$ and $g$ we can find quite sophisticated families of stationary and travelling localisations and many outcomes of the collisions and interactions between these localisations, an illustration is shown in Fig.~\ref{fig:complexcollision}. Thus the target could be, for example, to construct a $n$-binary full adder which is as compact in space and time as possible. 

The theoretical results reported show that by controlling just a few cells with Woronin bodies it is possible to drastically change dynamics of the automaton array. Third direction of future studies could be in implemented information processing in a single hypha. In such a hypothetical experimental setup input strings will be represented by arrays of illumination and outputs could be patterns of electrical activity recorded from the mycelium hypha resting on an electrode array. 

\section*{Acknowledgement}

AA, MT, HABW have received funding from the European Union's Horizon 2020 research and innovation programme FET OPEN ``Challenging current thinking'' under grant agreement No 858132. EG residency in UWE has been supported by funding from the Leverhulme Trust under the Visiting Research Professorship grant VP2-2018-001.

\bibliographystyle{plain}
\bibliography{references,fungalcomputerbib,grayscott}

\begin{thebibliography}{10}

\bibitem{adamatzkyspiking}
Andrew Adamatzky.
\newblock On spiking behaviour of oyster fungi {P}leurotus djamor.
\newblock {\em Scientific reports}, 7873, 2018.

\bibitem{adamatzky2018towards}
Andrew Adamatzky.
\newblock Towards fungal computer.
\newblock {\em Interface focus}, 8(6):20180029, 2018.

\bibitem{adamatzky2020boolean}
Andrew Adamatzky, Martin Tegelaar, Han~AB Wosten, Anna~L Powell, Alexander~E
  Beasley, and Richard Mayne.
\newblock On boolean gates in fungal colony.
\newblock {\em arXiv preprint arXiv:2002.09680}, 2020.

\bibitem{beasley2020memristive}
Alexander~E Beasley, Anna~L Powell, and Andrew Adamatzky.
\newblock Memristive properties of mushrooms.
\newblock {\em arXiv preprint arXiv:2002.06413}, 2020.

\bibitem{beck2013characterization}
Julia Beck and Frank Ebel.
\newblock Characterization of the major {W}oronin body protein hexa of the
  human pathogenic mold aspergillus fumigatus.
\newblock {\em International Journal of Medical Microbiology}, 303(2):90--97,
  2013.

\bibitem{berns1992optical}
Michael~W Berns, James~R Aist, William~H Wright, and Hong Liang.
\newblock Optical trapping in animal and fungal cells using a tunable,
  near-infrared titanium-sapphire laser.
\newblock {\em Experimental cell research}, 198(2):375--378, 1992.

\bibitem{bleichrodt2015switching}
Robert-Jan Bleichrodt, Marc Hulsman, Han~AB W{\"o}sten, and Marcel~JT Reinders.
\newblock Switching from a unicellular to multicellular organization in an
  aspergillus niger hypha.
\newblock {\em MBio}, 6(2):e00111--15, 2015.

\bibitem{bleichrodt2012hyphal}
Robert-Jan Bleichrodt, G~Jerre van Veluw, Brand Recter, Jun-ichi Maruyama,
  Katsuhiko Kitamoto, and Han~AB W{\"o}sten.
\newblock Hyphal heterogeneity in aspergillus oryzae is the result of dynamic
  closure of septa by {W}oronin bodies.
\newblock {\em Molecular microbiology}, 86(6):1334--1344, 2012.

\bibitem{bleichrodt2015selective}
Robert-Jan Bleichrodt, Arman Vinck, Nick~D Read, and Han~AB W{\"o}sten.
\newblock Selective transport between heterogeneous hyphal compartments via the
  plasma membrane lining septal walls of aspergillus niger.
\newblock {\em Fungal Genetics and Biology}, 82:193--200, 2015.

\bibitem{bonfante2009plants}
Paola Bonfante and Iulia-Andra Anca.
\newblock Plants, mycorrhizal fungi, and bacteria: a network of interactions.
\newblock {\em Annual review of microbiology}, 63:363--383, 2009.

\bibitem{carlile2001fungi}
Michael~John Carlile, Sarah~C Watkinson, and Graham~W Gooday.
\newblock {\em The fungi}.
\newblock Gulf Professional Publishing, 2001.

\bibitem{christensen1989view}
Martha Christensen.
\newblock A view of fungal ecology.
\newblock {\em Mycologia}, 81(1):1--19, 1989.

\bibitem{collinge1985woronin}
Annette~J Collinge and Paul Markham.
\newblock Woronin bodies rapidly plug septal pores of severedpenicillium
  chrysogenum hyphae.
\newblock {\em Experimental mycology}, 9(1):80--85, 1985.

\bibitem{cook2004universality}
Matthew Cook.
\newblock Universality in elementary cellular automata.
\newblock {\em Complex systems}, 15(1):1--40, 2004.

\bibitem{cooke1984ecology}
Roderic~C Cooke, Alan~DM Rayner, et~al.
\newblock {\em Ecology of saprotrophic fungi.}
\newblock Longman, 1984.

\bibitem{dai2011fomitiporia}
Yu-Cheng Dai and Bao-Kai Cui.
\newblock Fomitiporia ellipsoidea has the largest fruiting body among the
  fungi.
\newblock {\em Fungal biology}, 115(9):813--814, 2011.

\bibitem{deutsch1996zlib}
Peter Deutsch and Jean-Loup Gailly.
\newblock Zlib compressed data format specification version 3.3.
\newblock Technical report, 1996.

\bibitem{griffin1972ecology}
David~Michael Griffin et~al.
\newblock Ecology of soil fungi.
\newblock {\em Ecology of soil fungi.}, 1972.

\bibitem{held2008examining}
Marie Held, Clive Edwards, and Dan~V Nicolau.
\newblock Examining the behaviour of fungal cells in microconfined mazelike
  structures.
\newblock In {\em Imaging, Manipulation, and Analysis of Biomolecules, Cells,
  and Tissues VI}, volume 6859, page 68590U. International Society for Optics
  and Photonics, 2008.

\bibitem{held2009fungal}
Marie Held, Clive Edwards, and Dan~V Nicolau.
\newblock Fungal intelligence; or on the behaviour of microorganisms in
  confined micro-environments.
\newblock In {\em Journal of Physics: Conference Series}, volume 178, page
  012005. IOP Publishing, 2009.

\bibitem{howard1993design}
Paul~Glor Howard.
\newblock {\em The Design and Analysis of Efficient Lossless Data Compression
  Systems}.
\newblock PhD thesis, Citeseer, 1993.

\bibitem{jedd2000new}
Gregory Jedd and Nam-Hai Chua.
\newblock A new self-assembled peroxisomal vesicle required for efficient
  resealing of the plasma membrane.
\newblock {\em Nature cell biology}, 2(4):226--231, 2000.

\bibitem{leonhardt2017lah}
Yannik Leonhardt, Sara~Carina Kakoschke, Johannes Wagener, and Frank Ebel.
\newblock Lah is a transmembrane protein and requires spa10 for stable
  positioning of {W}oronin bodies at the septal pore of aspergillus fumigatus.
\newblock {\em Scientific reports}, 7:44179, 2017.

\bibitem{lew2005mass}
Roger~R Lew.
\newblock Mass flow and pressure-driven hyphal extension in neurospora crassa.
\newblock {\em Microbiology}, 151(8):2685--2692, 2005.

\bibitem{lindgren1990universal}
Kristian Lindgren and Mats~G. Nordahl.
\newblock Universal computation in simple one-dimensional cellular automata.
\newblock {\em Complex Systems}, 4(3):299--318, 1990.

\bibitem{martinez2003production}
Genaro~J. Mart{\'\i}nez, Harold~V McIntosh, and Juan C. Seck-Tuoh Mora.
\newblock Production of gliders by collisions in rule 110.
\newblock In {\em European Conference on Artificial Life}, pages 175--182.
  Springer, 2003.

\bibitem{martinez2006gliders}
Genaro~J. Mart{\'\i}nez, Harold~V McIntosh, and Juan C. Seck-Tuoh Mora.
\newblock Gliders in rule 110.
\newblock {\em International Journal of Unconventional Computing}, 2(1):1,
  2006.

\bibitem{mora2007rule}
Genaro~J. Mart{\'\i}nez, Juan C. Seck-Tuoh Mora, and Sergio~V.C. Vergara.
\newblock Rule 110 objects and other collision-based constructions.
\newblock {\em Journal of Cellular Automata}, 2:219--242, 2007.

\bibitem{maruyama2005three}
Jun-ichi Maruyama, Praveen~Rao Juvvadi, Kazutomo Ishi, and Katsuhiko Kitamoto.
\newblock Three-dimensional image analysis of plugging at the septal pore by
  woronin body during hypotonic shock inducing hyphal tip bursting in the
  filamentous fungus aspergillus oryzae.
\newblock {\em Biochemical and biophysical research communications},
  331(4):1081--1088, 2005.

\bibitem{momany2002mapping}
Michelle Momany, Elizabeth~A Richardson, Carole Van~Sickle, and Gregory Jedd.
\newblock Mapping {W}oronin body position in aspergillus nidulans.
\newblock {\em Mycologia}, 94(2):260--266, 2002.

\bibitem{moore1962fine}
Royall~T Moore and James~H McAlear.
\newblock Fine structure of mycota. 7. observations on septa of ascomycetes and
  basidiomycetes.
\newblock {\em American Journal of Botany}, 49(1):86--94, 1962.

\bibitem{neary2006p}
Turlough Neary and Damien Woods.
\newblock P-completeness of cellular automaton rule 110.
\newblock In {\em International Colloquium on Automata, Languages, and
  Programming}, pages 132--143. Springer, 2006.

\bibitem{ng2009tether}
Seng~Kah Ng, Fangfang Liu, Julian Lai, Wilson Low, and Gregory Jedd.
\newblock A tether for {W}oronin body inheritance is associated with
  evolutionary variation in organelle positioning.
\newblock {\em PLoS genetics}, 5(6), 2009.

\bibitem{olsson1995action}
S~Olsson and BS~Hansson.
\newblock Action potential-like activity found in fungal mycelia is sensitive
  to stimulation.
\newblock {\em Naturwissenschaften}, 82(1):30--31, 1995.

\bibitem{rayner1988fungal}
Alan~DM Rayner, Lynne Boddy, et~al.
\newblock {\em Fungal decomposition of wood. Its biology and ecology.}
\newblock John Wiley \& Sons Ltd., 1988.

\bibitem{roelofs1999png}
Greg Roelofs and Richard Koman.
\newblock {\em {PNG}: the definitive guide}.
\newblock O'Reilly \& Associates, Inc., 1999.

\bibitem{slayman1976action}
Clifford~L Slayman, W~Scott Long, and Dietrich Gradmann.
\newblock ``{A}ction potentials'' in {N}eurospora crassa, a mycelial fungus.
\newblock {\em Biochimica et Biophysica Acta (BBA)-Biomembranes},
  426(4):732--744, 1976.

\bibitem{smith1992fungus}
Myron~L Smith, Johann~N Bruhn, and James~B Anderson.
\newblock The fungus {A}rmillaria bulbosa is among the largest and oldest
  living organisms.
\newblock {\em Nature}, 356(6368):428, 1992.

\bibitem{soundararajan2004woronin}
Shanthi Soundararajan, Gregory Jedd, Xiaolei Li, Marilou Ramos-Pamplo{\~n}a,
  Nam~H Chua, and Naweed~I Naqvi.
\newblock Woronin body function in magnaporthe grisea is essential for
  efficient pathogenesis and for survival during nitrogen starvation stress.
\newblock {\em The Plant Cell}, 16(6):1564--1574, 2004.

\bibitem{steinberg2017woronin}
Gero Steinberg, Nicholas~J Harmer, Martin Schuster, and Sreedhar Kilaru.
\newblock {W}oronin body-based sealing of septal pores.
\newblock {\em Fungal Genetics and Biology}, 109:53--55, 2017.

\bibitem{tegelaar2020subpopulations}
Martin Tegelaar, Robert-Jan Bleichrodt, Benjamin Nitsche, Arthur~FJ Ram, and
  Han~AB W{\"o}sten.
\newblock Subpopulations of hyphae secrete proteins or resist heat stress in
  aspergillus oryzae colonies.
\newblock {\em Environmental microbiology}, 22(1):447--455, 2020.

\bibitem{tenney2000hex}
Karen Tenney, Ian Hunt, James Sweigard, June~I Pounder, Chadonna McClain,
  Emma~Jean Bowman, and Barry~J Bowman.
\newblock Hex-1, a gene unique to filamentous fungi, encodes the major protein
  of the woronin body and functions as a plug for septal pores.
\newblock {\em Fungal Genetics and Biology}, 31(3):205--217, 2000.

\bibitem{tey2005polarized}
Wei~Kiat Tey, Alison~J North, Jose~L Reyes, Yan~Fen Lu, and Gregory Jedd.
\newblock Polarized gene expression determines {W}oronin body formation at the
  leading edge of the fungal colony.
\newblock {\em Molecular biology of the cell}, 16(6):2651--2659, 2005.

\bibitem{trinci1974occlusion}
APJ Trinci and Annette~J Collinge.
\newblock Occlusion of the septal pores of damaged hyphae ofneurospora crassa
  by hexagonal crystals.
\newblock {\em Protoplasma}, 80(1-3):57--67, 1974.

\bibitem{wergin1973development}
William~P Wergin.
\newblock Development of {W}oronin bodies from microbodies infusarium oxysporum
  f. sp. lycopersici.
\newblock {\em Protoplasma}, 76(2):249--260, 1973.

\bibitem{wolfram1984universality}
Stephen Wolfram.
\newblock Universality and complexity in cellular automata.
\newblock {\em Physica D: Nonlinear Phenomena}, 10(1-2):1--35, 1984.

\bibitem{wolfram1994cellular}
Stephen Wolfram.
\newblock {\em Cellular automata and complexity: collected papers}.
\newblock Addison-Wesley Pub. Co., 1994.

\bibitem{ziv1977universal}
Jacob Ziv and Abraham Lempel.
\newblock A universal algorithm for sequential data compression.
\newblock {\em IEEE Transactions on information theory}, 23(3):337--343, 1977.

\end{thebibliography}
\end{document}